\documentclass[12pt]{article}
\usepackage{fullpage,amsfonts,amssymb,pstricks,pst-node,epsfig}

\begin{document}

\newtheorem{theorem}{Theorem}

\newcommand{\be}{\begin{eqnarray}}
\newcommand{\ee}{\end{eqnarray}}
\newcommand{\ba}{\begin{array}}
\newcommand{\ea}{\end{array}}
\newcommand{\nn}{\nonumber}

\def\s{\scriptstyle}
\def\ss{\scriptscriptstyle}
\def\ds{\displaystyle}
\def\p{\partial}

\renewcommand{\topfraction}{0.85}
\renewcommand{\textfraction}{0.1}
\psset{dash=2mm 1mm,dotsep=0.5mm}

\begin{titlepage}
\setcounter{page}{0}
\thispagestyle{empty}

\begin{flushright}
May 10, 2001
\end{flushright}
\vskip 0.25truein

\begin{center}
\LARGE \bf The Generalized Gell-Mann--Low Theorem \\[1.5mm]
for Relativistic Bound States{\Large\footnote{This work was supported 
by CIC--UMSNH and Conacyt grant 32729--E.}} \\[0.35truein]
\large \rm Axel Weber\footnotemark[2] and 
Norbert E. Ligterink\footnotemark[3] \\[0.25truein]
\normalsize \it \footnotemark[2]Instituto de F\'\i sica y Matem\'aticas, 
Universidad Michoacana de San Nicol\'as de Hidalgo \\
Edificio C-3 Cd.\ Universitaria, A. Postal 2--82 \\
58040 Morelia, Michoac\'an, Mexico \\
e--mail:\/ {\sf axel@io.ifm.umich.mx} \\[0.2truein]
\footnotemark[3]ECT\renewcommand
{\thefootnote}{\,\fnsymbol{footnote}}\footnotemark[1] \\
Villa Tambosi, Strada delle Tabarelle 286 \\
I--38050 Villazzano (Trento), Italy \\
e--mail:\/ {\sf ligterin@ect.it} 
\end{center}
\vskip 0.6truein

\noindent
{\bf Abstract.} The recently established generalized Gell-Mann--Low theorem
is applied in lowest perturbative order to bound--state 
calculations in a simple scalar field theory with cubic couplings. 
The approach via the generalized Gell-Mann--Low Theorem retains, while 
being fully relativistic, many of the desirable features of the quantum 
mechanical approaches to bound states. In particular, no abnormal or 
unphysical solutions are found in the model under consideration. 
Both the non-relativistic and one--body limits are straightforward and 
consistent. The results for the spectrum are compared to those of the 
Bethe--Salpeter equation (in the ladder approximation) and related
equations. 

\vfill

\end{titlepage}
\setcounter{footnote}{3}

\section{Introduction}

The generalization of the Gell-Mann--Low or adiabatic theorem \cite{GML51}
(see also \cite{FW71}) was established recently \cite{Web99} by one of 
the authors, with the intention of making a certain
class of ``non--perturbative'' phenomena accessible to perturbative methods.
The emergence of bound states is a perfect paradigm for such
phenomena: from a Lagrangian field 
theory point of view, bound states are characterized by poles in the
S--matrix and as such can only be generated by summing an infinite set of
Feynman diagrams \cite{IZ80}. The most traditional method is to choose a 
particular (finite) set of diagrams contributing to the four--point function,
and to iterate these diagrams to all orders via an integral equation,
as, for example, the Bethe--Salpeter equation \cite{GML51,BS51} (for a 
review see Ref.\ \cite{Nak69}). In that case all two--particle reducible 
products of the initial set are included in the final 
result. Therefore the initial set itself should contain only two--particle
irreducible diagrams. In most calculations only the one--particle 
exchange diagram is taken into account in the kernel of the integral equation. 
The iteration of this kernel through the solution of the integral equation 
leads to the set of so--called ladder diagrams, and the corresponding
approximation to the four--point function is known as the ladder 
approximation.

The Bethe--Salpeter equation, or rather the Bethe--Salpeter approach, has 
many problems \cite{IZ80,Nak69}, most of which
are related to the appearance of a relative--time coordinate. The 
covariant, Lagrangian, or space--time formulation of field theory was 
originally designed for scattering problems, where the interaction distances
are short and consequently the interaction time is short as well. However,
for a bound state, the ``interaction time'' is infinite, and for massless 
exchange particles the interaction range is large. For such problems the 
Hamiltonian, time--independent
approach is better suited, but difficult to implement consistently.

Since the formulation of the Bethe--Salpeter equation, there has been a
wealth of alternative proposals for bound--state equations in quantum field 
theory. In the beginning, despite the drawbacks mentioned above, they all 
started from the Bethe--Salpeter equation, with the relative time eliminated 
in one way or another. These equations are generically known as quasipotential 
equations \cite{Tod71}, with the Blankenbecler--Sugar 
equation \cite{BS66,LT63} and the Gross or spectator equation \cite{Gro69}
as well--known examples. 
However, even though these equations have to satisfy some unitarity conditions,
much arbitrariness is left, and results depend on the way the 
four--dimensional equations are reduced to three--dimensional ones 
\cite{PW98}.

It is important to notice that all these approaches are rooted
in the Bethe--Salpeter equation and therefore in Lagrangian perturbation
theory. For the description of physical states, and in particular bound states,
as mentioned before, the Hamiltonian formalism is preferable over the 
Lagrangian one, the tool of choice being the time--independent
Schr\"odinger equation. Indeed, it is clear that in a Hamiltonian approach a 
relative--time coordinate has no place, thus eliminating one of the
major problems of the Lagrangian approach right from the start. 

More recently, other routes to bound--state equations have been explored.
Here we just mention a few. Firstly, there is the Feynman--Schwinger 
Representation (FSR) approach \cite{ST93}--\cite{NT96}, which starts from the 
path--integral representation of the four--point function. The integral
is performed partly analytically and partly with Monte Carlo techniques
in Euclidean space, hence the FSR is closely related to the Lagrangian 
approach. However, most problems associated with the Bethe--Salpeter equation 
are absent, or not yet discovered, as it is hard to recover excited states in 
this approach. 

Secondly, there are a number of light--front field theory approaches
\cite{BPP98}--\cite{Han00}, which 
either work in discrete momentum space, or at low orders
in a Fock--space expansion, but cover a wide range of approaches from rather
traditional to innovative. In all these Hamiltonian approaches,
(approximations to) physical states are calculated. 

Thirdly, a few investigations have been made using the Haag expansion, or 
N--quantum approach \cite{GRS95}, but, probably because of computational 
difficulties, it has never been successful. 
Fourthly, one of the authors has recently explored the same problem
in an ordinary equal--time Hamiltonian formulation with a Fock--space 
truncation \cite{LB01}. Self--energy effects have been included, which had been
ignored in earlier work.
Finally, the other author has applied Regge theory to extract bound state
energies from the leading Regge trajectories as calculated by one--loop
renormalization group improved Lagrangian perturbation theory \cite{WLSD00}.

However, few of these approaches formulate a perturbative approximation to
an effective Hamiltonian, which can then form the basis of a bound--state 
equation in a given particle sector.
This and the beforegoing considerations have been the major motivation for the
generalization of the Gell-Mann--Low theorem, which may be considered to
be a particularly efficient formulation of Hamiltonian perturbation theory.
Other non--perturbative phenomena like spontaneous symmetry
breaking and vacuum condensation are hoped to be describable by the same
method. In each case, the application of the Gell-Mann--Low theorem leads
to a series of perturbative approximations to a certain equation. In analogy
with the integral equations of the traditional Lagrangian approach, it is the 
solution of this equation (or any of its approximations) which leads to the 
wanted non--perturbative information. In the case of the bound--state problem,
this equation is nothing but the time--independent Schr\"odinger
equation, where the potential part is given in the form of a perturbative 
series.

Unfortunately, in this Hamiltonian approach there
is also a counterpart to the arbitrariness in the Lagrangian approach
to bound states: different mappings to the unperturbed subspace that appears
in the generalization of the Gell-Mann--Low theorem,
lead to non--equivalent Schr\"odinger equations at finite orders of the
perturbative expansion. A future discussion of 
this issue will take properties like renormalizability, and Lorentz and gauge 
invariance at finite orders of the perturbative expansion into account, but
it will also have to consider applications to simple models as in the 
present contribution. For the time being, we will focus on the particular 
choice discussed in Ref.\ \cite{Web99}, which has somewhat unique properties.

In this paper we will demonstrate the applicability of the 
generalized Gell-Mann--Low theorem
to bound--state problems with the help of a very simple model, and compare with
the Bethe--Salpeter and related approaches, arguing the superiority of the 
former method
over the latter. The problem of choice is the calculation of bound states of
two distinguishable scalar constituents interacting through the exchange
of a third scalar particle. In the case of a massless exchange particle,
and in the ladder approximation to the Bethe--Salpeter equation, this model
is known as the Wick--Cutkosky model \cite{Nak69,WC54}. 

In the following section, we will derive the 
effective Schr\"odinger equation for this
scalar model from the generalized Gell-Mann--Low theorem to lowest 
non--trivial order, emphasizing the diagrammatic representation of the
effective Hamiltonian. The resulting diagrams are similar to Goldstone 
diagrams \cite{Gol57}, but, unlike the latter, in general they cannot be 
combined into on--shell Feynman diagrams. 
The third section presents an analytical study of the
non--relativistic and one--body limits, demonstrating the consistency of the
formalism in these particular cases.
In the fourth section, we solve the effective Schr\"odinger 
equation numerically, establishing that all solutions are physical and
confirming the analytical results of the preceding section. Finally, the last 
section presents a critical discussion of the results and a comparison with 
the Bethe--Salpeter equation and other approaches.

\section{The Schr\"odinger Equation}

The new method we propose and explore here for the solution of the bound--state
problem in quantum field theory, is a generalization of the Gell-Mann--Low
theorem \cite{Web99}. The essential statement of the generalized Gell-Mann--Low
theorem (GGL for short) is that the adiabatic evolution operator maps linear
subspaces invariant under the free Hamiltonian $H_0$ to linear subspaces that
are invariant under the {\em full}\/ Hamiltonian $H = H_0 + H_I$. The full 
Hamiltonian is then diagonalizable in that subspace and one need not consider
the eigenvalue problem in the whole Fock space. The mapping between the two
linear subspaces considered in Refs.\ \cite{Web99} is such that when followed
by the orthogonal projection from the $H$--invariant subspace back to the 
$H_0$--invariant subspace, it gives the identical map in the latter. In order
to assure this normalization, the adiabatic evolution operator has to be
accompanied by a mapping within the $H_0$--invariant subspace, and the 
combination of these two maps has been called the Bloch--Wilson operator.

The proof presented in Ref.\ \cite{Web99} rests on the existence of a 
perturbative series for the Bloch--Wilson operator, or more precisely on the 
existence of its adiabatic limit. It should not be taken for 
granted, but rather depends on a sensible choice for
the $H_0$--invariant subspace. In practice, one will only calculate to a 
certain perturbative order, and the existence of the limit is then only 
guaranteed (or needed) up to that order. The explicit calculation to be 
performed in the present work will demonstrate that a reasonable approximation
can be achieved by this procedure to lowest order, and that indeed a sensible
and natural choice (to this order) for the $H_0$--invariant subspace exists. 
It should also be noted that the Bloch--Wilson operator establishes
a similarity transformation between the two linear subspaces so 
that the problem of diagonalizing the Hamiltonian $H$ in the $H$--invariant 
subspace can be transported back to the $H_0$--invariant subspace. The image 
of $H$ under the similarity transformation in that subspace is $H_{BW}$, the 
Bloch--Wilson Hamiltonian. Below, $H_{BW}$ will be calculated to lowest
non--trivial order for the problem at hand.

We will now introduce our notations and also restate the basic
formulas (for their derivation and discussion see Ref.\ 
\cite{Web99}). The adiabatic evolution operator $U_\epsilon$ is given by
the series expansion
\be
U_{\epsilon} = \sum_{n=0}^\infty \frac{(-i)^n}{n!}
\int_{-\infty}^0 dt_1 \cdots \int_{-\infty}^0 dt_n \, e^{-\epsilon (|t_1| +
\ldots + |t_n|)} \, T [ H_I (t_1) \cdots H_I (t_n) ] \:,
\ee
where
\be
H_I(t) = e^{i H_0 t} H_I \, e^{-i H_0 t} \:.
\ee
The Bloch--Wilson operator $U_{BW}$ is defined by the adiabatic limit
\be
U_{BW} = \lim_{\epsilon \to 0} \, U_\epsilon (P_0 U_\epsilon P_0)^{-1} \:,
\label{BWopgen}
\ee
$P_0$ being the orthogonal projection to the $H_0$--invariant subspace
$\Omega_0$. Granted the existence of $U_{BW}$, the Bloch--Wilson 
Hamiltonian is simply
\be
H_{BW} = P_0 H U_{BW} \:. \label{BWhamgen}
\ee

The scalar model we consider in the following consists of three scalar 
particles, two of them carrying some charge and one neutral, with masses
$m_A$, $m_B$ and $\mu$, respectively. The two charged particles are
coupled to the neutral one via a charge--conserving cubic coupling with
strength $g$. The coupling constant $g$ has the dimension of a mass,
and for our purposes we might as well have chosen two different constants
$g_A$ and $g_B$ for the coupling to the two charged fields. The Hamiltonian
of the model is
\be
H &=& H_0 + H_I \:, \nn \\[2mm]
H_0 &=& \int d^3 x \Bigg[ \phi_A^\dagger ({\bf x}) (m_A^2 - \nabla^2) 
\phi_A ({\bf x}) + \phi_B^\dagger ({\bf x}) (m_B^2 - \nabla^2) 
\phi_B ({\bf x}) \nn \\
& & \left. {}+ \frac{1}{2} \varphi ({\bf x})
(\mu^2 - \nabla^2) \varphi ({\bf x}) \right] \:, \nn \\[2mm]
H_I &=& \mbox{\bf :} \int d^3 x \left[ g \phi_A^\dagger ({\bf x}) 
\phi_A ({\bf x}) \varphi ({\bf x}) + g \phi_B^\dagger ({\bf x}) 
\phi_B ({\bf x}) \varphi ({\bf x}) \right] \mbox{\bf :} \:, \label{model}
\ee
where we have normal--ordered the interaction part, mainly for reasons of
presentation. It is well--known that the perturbative vacuum is unstable in 
this model field theory. However, perturbation theory around this vacuum is 
well--defined order by order,  and no instability appears as long 
as the number of (perturbative) constituent particles in any intermediate 
state is finite. This model has been much used in the past to test 
bound--state equations. In particular, it has played an important r\^ole for 
the Bethe--Salpeter equation, 
where the ladder approximation in the special case $\mu = 0$ is
known as the Wick--Cutkosky model \cite{WC54}. The bound states usually 
considered are the ones containing one $A$-- and one $B$--particle as
constituents. The model is then the simplest one can possibly think of.

The Fock space will be built up as usual by repeated application of the
creation operators to the vacuum $| \Omega \rangle$, $\langle \Omega | \Omega
\rangle = 1$. We will normalize the creation and annihilation operators in
momentum space in such a way that
\be
[ a_i ({\bf p}), a_i^\dagger ({\bf p}') ] = (2 \pi)^3 
\delta ({\bf p} - {\bf p}') \label{comm}
\ee
for all three particles (labelled by the index $i$). This normalization is
of course not relativistically invariant, but it turns out to be convenient,
and is closely related to the quantum--mechanical wave functions.
Besides, Lorentz invariance is not manifest in a Hamiltonian approach anyway.

We can now define the  $H_0$--invariant subspace $\Omega_0$ we will use for the
application of the GGL. Considering that we want to determine bound states of
an $A$--and a $B$--particle which will contain components with different
momenta of these particles, a natural choice is
\be
\Omega_0 = \mbox{span} \left. \left\{ | {\bf p}_A, {\bf p}_B \rangle
= a_A^\dagger ({\bf p}_A) a_B^\dagger ({\bf p}_B) | \Omega \rangle \right|
{\bf p}_A, {\bf p}_B \in \mathbb{R}^3 \right\} \:, \label{omtwop}
\ee
and the corresponding orthogonal projector reads
\be
P_0 = \int \frac{d^3 p_A}{(2 \pi)^3} \frac{d^3 p_B}{(2 \pi)^3}
| {\bf p}_A, {\bf p}_B \rangle \langle {\bf p}_A, {\bf p}_B | \:.
\ee
We could have restricted $\Omega_0$ to states with fixed total momentum
${\bf P} = {\bf p}_A + {\bf p}_B$, but it is more instructive to see how
this constraint arises from momentum conservation.

The rest of this section will consist in calculating and interpreting the 
Bloch--Wilson Hamiltonian $H_{BW}$ to lowest non--trivial order. In
Eqs.\ (\ref{BWopgen}) and (\ref{BWhamgen}), $H_{BW}$ was
defined as the adiabatic limit $\epsilon \to 0$ of the operator
\be
& & P_0 H U_\epsilon (P_0 U_\epsilon P_0)^{-1} \nn \\
&=& P_0 H_0 U_\epsilon (P_0 U_\epsilon P_0)^{-1} +
P_0 H_I U_\epsilon (P_0 U_\epsilon P_0)^{-1} \nn \\
&=& H_0 P_0 + P_0 H_I U_\epsilon (P_0 U_\epsilon P_0)^{-1} \:,
\ee
where in the last step we have made use of the fact that $\Omega_0$ is a
direct sum of eigenspaces of $H_0$, hence $P_0 H_0 = H_0 P_0$, 
and of the fact that one can 
insert an operator $P_0$ between $U_\epsilon$ and $(P_0 U_\epsilon P_0)^{-1}$.
To first order in $H_I$, one needs to keep only the trivial term for 
$U_\epsilon$, i.e.\ $U_\epsilon = {\bf 1} + {\cal O} (H_I)$. But since $H_I$ 
necessarily changes the number of particles, one has $P_0 H_I P_0 = 0$, and 
$H_{BW}$ reduces to the non--interacting part, $H_{BW} = H_0 P_0$.

Consequently, for the lowest non--trivial order one has to expand $U_\epsilon$
to first order in $H_I$, i.e.,
\be
U_\epsilon = {\bf 1} - i \int_{-\infty}^0 dt \, e^{- \epsilon |t|}
H_I (t) + {\cal O} (H_I^2) \:.
\ee
Using $P_0 H_I P_0 = 0$ as before, the Bloch--Wilson Hamiltonian to lowest
non--trivial order is given by
\be
H_{BW} = H_0 P_0 - i \int_{-\infty}^0 dt \, e^{- \epsilon |t|} P_0
H_I (0) H_I (t) P_0 + {\cal O} (H_I^4) \:, \label{HBW}
\ee
where the limit $\epsilon \to 0$ is understood.

\begin{figure}
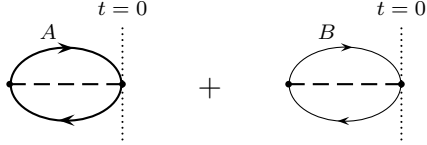

\begin{center} 
\pspicture[0.4](-0.25,0)(3,1.5)
\psellipse(1.25,0.75)(0.75,0.5)
\psline{->}(1.25,1.23)(1.35,1.23)
\uput[135](1.25,1.23){\mbox{\scriptsize $A$}}
\psline{->}(1.25,0.27)(1.15,0.27)
\psdots(0.5,0.75)(2,0.75)
\psline[linestyle=dashed](0.5,0.75)(2,0.75)
\psline[linestyle=dotted](2,0)(2,1.5)
\uput[90](2,1.5){\mbox{\scriptsize $t=0$}}
\endpspicture + \pspicture[0.4](-0.25,0)(3,1.5)
\psellipse[linewidth=.4pt](1.25,0.75)(0.75,0.5)
\psline[linewidth=.4pt]{->}(1.25,1.24)(1.35,1.24)
\uput[135](1.25,1.23){\mbox{\scriptsize $B$}}
\psline[linewidth=.4pt]{->}(1.25,0.26)(1.15,0.26)
\psdots(0.5,0.75)(2,0.75)
\psline[linestyle=dashed](0.5,0.75)(2,0.75)
\psline[linestyle=dotted](2,0)(2,1.5)
\uput[90](2,1.5){\mbox{\scriptsize $t=0$}}
\endpspicture
\end{center}
\caption{The second--order contributions to the vacuum energy according to
Eq.\ (\ref{vaccor}). The propagators of the charged $A$-- and $B$--particles
are represented by thick and thin lines, respectively, while for the neutral 
particle we have drawn broken lines. Integration over $t$ from $-\infty$ to 
$0$ for the vertex to the left and integration over the spatial position 
$\bf x$ for {\em both}\/ vertices is understood. \label{vacfig}}
\end{figure}

We could now go on and directly determine an explicit expression for $H_{BW}$
to this order, but it is conceptually clearer (and in principle necessary)
to consider the zero-- and one--particle sectors first. A posteriori we will
see how an analysis of the diagrams contributing to the $H_{BW}$ of Eq.\
(\ref{HBW})
yields the same information. To begin with, we study the vacuum sector to the
same order in the coupling constant $g$. The Hamiltonian in Eq.\ (\ref{HBW}) 
then remains 
unchanged except for the replacement of $P_0$ with the vacuum projector
\be
P_\Omega = | \Omega \rangle \langle \Omega | \:.
\ee
We can simplify Eq.\ (\ref{HBW}) to
\be
H_{BW} = \left( E_0 - i \int_{-\infty}^0 dt \, e^{- \epsilon |t|}
\langle \Omega | T [ H_I (0) H_I (t) ] | \Omega \rangle 
\right) P_\Omega + {\cal O} (H_I^4) \:, \label{HBWvac}
\ee
where $E_0 = \langle \Omega | H_0 | \Omega \rangle$ is the vacuum energy, due
to the fact that we have not normal--ordered $H_0$ (there would be an 
additional contribution to second order in $g$, had we not normal-ordered
$H_I$). By use of Wick's theorem, the matrix element in the 
second--order contribution to $H_{BW}$ can be written in terms of Feynman 
propagators in position space as
\be
\lefteqn{\langle \Omega | T [
H_I (0) H_I (t) ] | \Omega \rangle} \nn \\
&=& g^2 \int d^3 x \, d^3 x'
\Delta_F^A (0 - t, {\bf x} - {\bf x}') \Delta_F^A (t - 0, {\bf x}' - {\bf x})
\Delta_F (0 - t, {\bf x} - {\bf x}') \nn \\
& & {}+ g^2 \int d^3 x \, d^3 x' 
\Delta_F^B (0 - t, {\bf x} - {\bf x}') \Delta_F^B (t - 0, {\bf x}' - {\bf x})
\Delta_F (0 - t, {\bf x} - {\bf x}') \:, \label{vaccor}
\ee
where
\be
\Delta_F (t, {\bf x}) &=& i \int \frac{d^4 k}{(2 \pi)^4} \frac{e^{-i k_0 t
+ i {\bf k} \cdot {\bf x}}}{k^2 - \mu^2 + i \epsilon}  \label{feynmcov} \\
&=& \int \frac{d^3 k}{(2 \pi)^3 2 \omega_{\bf k}} \left[ \theta(t)
 e^{- i (\omega_{\bf k} t - {\bf k} \cdot {\bf x})} + \theta(-t) 
e^{i (\omega_{\bf k} t - {\bf k} \cdot {\bf x})} \right]  \:,
\label{feynmnon}
\ee
with
\be
\omega_{\bf k} = \sqrt{\mu^2 + {\bf k}^2} \:,
\ee
and analogously for the $A$-- and $B$--particles. Eq.\ (\ref{vaccor})
is represented diagrammatically in Fig.\ \ref{vacfig}.

Using the non--covariant representation Eq.\ (\ref{feynmnon}) of the Feynman 
propagators and performing the integrations over $t$, ${\bf x}$, and
${\bf x}'$, the second--order contribution in Eq.\ (\ref{HBWvac}) is cast in 
the alternative form
\be
E_\Omega - E_0 &=& - i \int_{-\infty}^0 dt \, e^{- \epsilon |t|} 
\langle \Omega | T [
H_I (0) H_I (t) ] | \Omega \rangle \nn \\
&=& - g^2 \left( \int \frac{d^3 k}{(2 \pi)^3 2 
\omega^A_{\bf k}} \, \frac{d^3 k'}{(2 \pi)^3 2 \omega^A_{{\bf k}'}} \, \frac{1}
{2 \omega_{{\bf k} + {\bf k}'} (\omega^A_{\bf k} + \omega^A_{{\bf k}'}
+ \omega_{{\bf k} + {\bf k}'})} \right. \nn \\
& & \left. {}+ \int \frac{d^3 k}{(2 \pi)^3 2 \omega^B_{\bf k}} \, 
\frac{d^3 k'}{(2 \pi)^3 2 \omega^B_{{\bf k}'}} \, \frac{1}
{2 \omega_{{\bf k} + {\bf k}'} (\omega^B_{\bf k} + \omega^B_{{\bf k}'}
+ \omega_{{\bf k} + {\bf k}'})} \right) (2 \pi)^3 \delta^{(3)} (0) \:.
\hspace{5mm} \label{vaccorr}
\ee
Of course, this expression is not very meaningful as it stands. 
Some regularization procedure would have to be employed 
to render it finite. On the other hand, what will be of interest is only the 
{\em difference}\/ in energy between the vacuum and other states. 
Since the expression Eq.\ (\ref{vaccorr}) always appears as the 
result of evaluating the diagrams of Fig.\ \ref{vacfig}, it will
always be easy to identify and subtract it. The same expression results
in usual time--independent Schr\"odinger perturbation theory and also in
covariant Lagrangian perturbation theory to second order.

Having treated the vacuum sector in some detail, we now go on to discuss the
one--particle states, taking as an example the momentum eigenstates of one
$A$--particle. The natural choice for the corresponding $H_0$--invariant
subspace is $\Omega_A$, the span of all these states, and the corresponding
orthogonal projector is
\be
P_A = \int \frac{d^3 p_A}{(2 \pi)^3} | {\bf p}_A \rangle \langle {\bf p}_A | 
\label{onepproj}
\:.
\ee

The corresponding Bloch--Wilson Hamiltonian to lowest non--trivial order is
again given by Eq.\ (\ref{HBW}), with $P_A$ replacing $P_0$, and the relevant
matrix elements of the second--order term are
\be
\lefteqn{\langle {\bf p}_A | T [
H_I (0) H_I (t) ] | {\bf p}_A' \rangle} \nn \\
&=& g^2 \int d^3 x \, d^3 x'
\Delta_F^A (0 - t, {\bf x} - {\bf x}') \Delta_F^A (t - 0, {\bf x}' - {\bf x})
\Delta_F (0 - t, {\bf x} - {\bf x}') 
(2 \pi)^3 \delta ({\bf p}_A - {\bf p}_A') \nn \\
& & {}+ g^2 \int d^3 x \, d^3 x' 
\Delta_F^B (0 - t, {\bf x} - {\bf x}') \Delta_F^B (t - 0, {\bf x}' - {\bf x})
\Delta_F (0 - t, {\bf x} - {\bf x}') 
(2 \pi)^3 \delta ({\bf p}_A - {\bf p}_A') \nn \\
& & {}+ g^2 \int d^3 x \, d^3 x' \psi_{{\bf p}_A}^{A \, \ast} (0, {\bf x}) 
\Delta_F^A (0 - t, {\bf x} - {\bf x}') \Delta_F (0 - t, {\bf x} - {\bf x}')
\psi_{{\bf p}_A'}^A (t, {\bf x}') \nn \\
& & {}+ g^2 \int d^3 x \, d^3 x' \psi_{{\bf p}_A}^{A \, \ast} (t, {\bf x}') 
\Delta_F^A (t - 0, {\bf x}' - {\bf x}) \Delta_F (t - 0, {\bf x}' - {\bf x})
\psi_{{\bf p}_A'}^A (0, {\bf x}) \:, \label{onepcor}
\ee
in terms of Feynman propagators, where the wave functions 
$\psi_{\bf p}^A (t, {\bf x})$ are given by
\be
\psi_{\bf p}^A (t, {\bf x}) = \frac{e^{-i \omega_{\bf p}^A t +
i {\bf p} \cdot {\bf x}}}{\sqrt{2 \omega_{\bf p}^A}} \:, \label{wavef}
\ee
minding the normalization of the one--particle states implicit in 
Eq.\ (\ref{comm}). To arrive at Eq.\ (\ref{onepcor}), Wick's theorem has
been used, including contractions like
\be
\rnode{A}{a_A ({\bf p})} \, \rnode{B}{\phi_A^\dagger (t, {\bf x})} = 
\psi^{A \, \ast}_{\bf p} (t, {\bf x}) \:, \quad 
\ncbar[linewidth=.5pt,nodesep=3pt,arm=5pt,angle=-90]{A}{B}
\rnode{C}{\phi_A (t, {\bf x})} \, \rnode{D}{a_A^\dagger ({\bf p})} = 
\psi^A_{\bf p} (t, {\bf x}) \:.
\ncbar[linewidth=.5pt,nodesep=3pt,arm=5pt,angle=-90]{C}{D}
\ee
\vspace{-3mm}

\begin{figure}
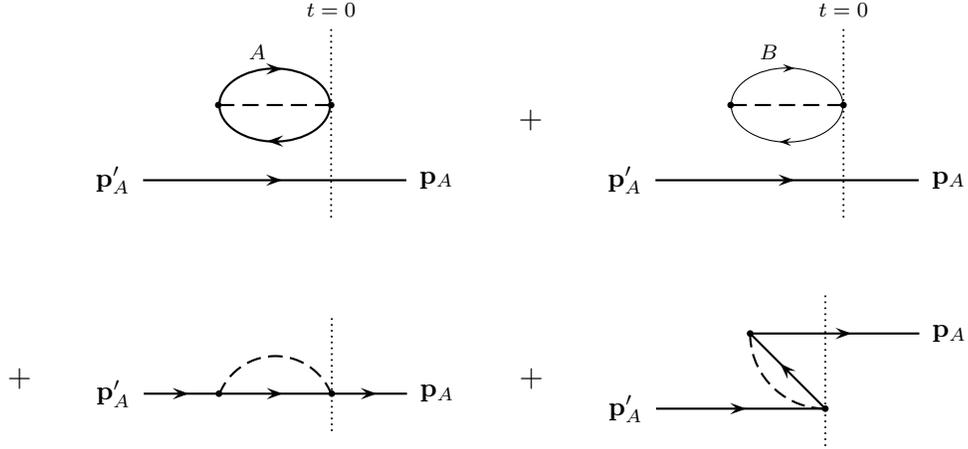

\begin{center} 
\begin{eqnarray*}
& & \pspicture[0.4](-1.4,1)(5.4,4)
\psellipse(2.25,2.5)(0.75,0.5)
\psline{->}(2.25,2.98)(2.35,2.98)
\uput[135](2.25,2.98){\mbox{\scriptsize $A$}}
\psline{->}(2.25,2.02)(2.15,2.02)
\psdots(1.5,2.5)(3,2.5)
\psline[linestyle=dashed](1.5,2.5)(3,2.5)
\psline(2.25,1.5)(4,1.5)
\uput[0](4,1.5){\mbox{\footnotesize ${\bf p}_A$}}
\psline{->}(0.5,1.5)(2.35,1.5)
\uput[180](0.5,1.5){\mbox{\footnotesize ${\bf p}_A'$}}
\psline[linestyle=dotted](3,1)(3,3.5)
\uput[90](3,3.5){\mbox{\scriptsize $t=0$}}
\endpspicture + \pspicture[0.4](-0.9,1)(5.4,4)
\psellipse[linewidth=.4pt](2.25,2.5)(0.75,0.5)
\psline[linewidth=.4pt]{->}(2.25,2.99)(2.35,2.99)
\uput[135](2.25,2.98){\mbox{\scriptsize $B$}}
\psline[linewidth=.4pt]{->}(2.25,2.01)(2.15,2.01)
\psdots(1.5,2.5)(3,2.5)
\psline[linestyle=dashed](1.5,2.5)(3,2.5)
\psline(2.25,1.5)(4,1.5)
\uput[0](4,1.5){\mbox{\footnotesize ${\bf p}_A$}}
\psline{->}(0.5,1.5)(2.35,1.5)
\uput[180](0.5,1.5){\mbox{\footnotesize ${\bf p}_A'$}}
\psline[linestyle=dotted](3,1)(3,3.5)
\uput[90](3,3.5){\mbox{\scriptsize $t=0$}}
\endpspicture \\[9mm]
& & {} + \pspicture[0.4](-0.9,1.5)(5.4,3)
\psline(3.5,2)(4,2)
\uput[0](4,2){\mbox{\footnotesize ${\bf p}_A$}}
\psline{->}(2.25,2)(3.6,2)
\psline{->}(1,2)(2.35,2)
\psline{->}(0.5,2)(1.1,2)
\uput[180](0.5,2){\mbox{\footnotesize ${\bf p}_A'$}}
\psdots(1.5,2)(3,2)
\psarc[linestyle=dashed](2.25,1.7){0.8}{20}{160}
\psline[linestyle=dotted](3,1.5)(3,3)
\endpspicture + \pspicture[0.4](-0.9,1)(5.4,3)
\psline(3,2.5)(4,2.5)
\uput[0](4,2.5){\mbox{\footnotesize ${\bf p}_A$}}
\psline{->}(1.75,2.5)(3.1,2.5)
\psline(2.25,2)(1.75,2.5)
\psline{->}(2.75,1.5)(2.15,2.1)
\psline(1.6,1.5)(2.75,1.5)
\psline{->}(0.5,1.5)(1.7,1.5)
\uput[180](0.5,1.5){\mbox{\footnotesize ${\bf p}_A'$}}
\psdots(2.75,1.5)(1.75,2.5)
\psarc[linestyle=dashed](2.75,2.5){1}{-180}{-90}
\psline[linestyle=dotted](2.75,1)(2.75,3)
\endpspicture
\end{eqnarray*}
\end{center}
\caption{The second--order contributions to the one--particle states 
representing Eq.\ (\ref{onepcor}). The external lines with momentum
labels correspond to the wave functions Eq.\ (\ref{wavef}), representing the 
free states. \label{onepfig}}
\end{figure}

The four terms appearing in Eq.\ (\ref{onepcor}) are represented
diagrammatically in Fig.\ \ref{onepfig}.
The first two terms in Eq.\ (\ref{onepcor}) are diagonal in the basis 
$\{ | {\bf p}_A \rangle \}$,
and a comparison with Eq.\ (\ref{vaccor}) shows that, together with the
contribution $E_0 (2 \pi)^3 \delta ({\bf p}_A - {\bf p}_A')$ to order zero,
they just reproduce the vacuum energy $E_\Omega$ 
calculated before. On a diagrammatic level, these contributions precisely
correspond to the unlinked diagrams in Fig.\ \ref{onepfig}. An unlinked
diagram by definition contains a part that is not connected (or linked) to 
any external line.

\begin{figure}
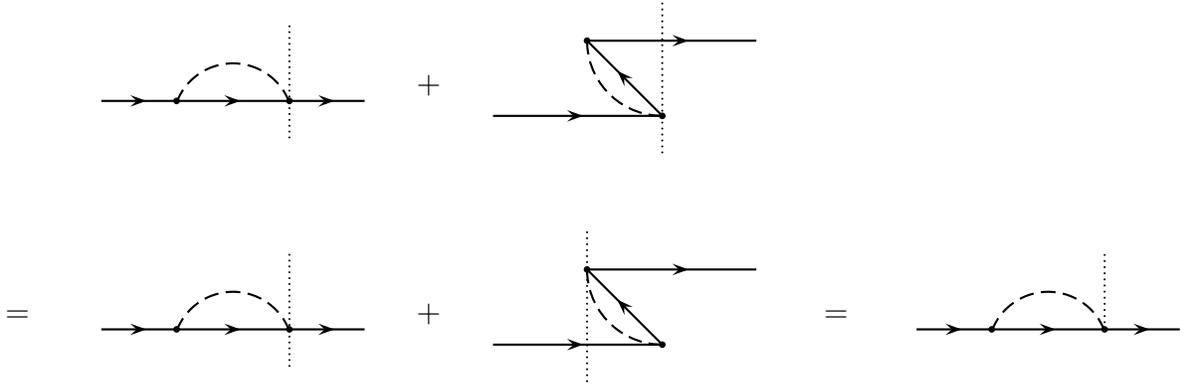

\begin{center}
\begin{eqnarray*}
& & \pspicture[0.4](-0.1,0)(4.6,1.5)
\psline(3.5,0.5)(4,0.5)
\psline{->}(2.25,0.5)(3.6,0.5)
\psline{->}(1,0.5)(2.35,0.5)
\psline{->}(0.5,0.5)(1.1,0.5)
\psdots(1.5,0.5)(3,0.5)
\psarc[linestyle=dashed](2.25,0.2){0.8}{20}{160}
\psline[linestyle=dotted](3,0)(3,1.5)
\endpspicture + \pspicture[0.4](-0.1,0)(4.6,2)
\psline(3,1.5)(4,1.5)
\psline{->}(1.75,1.5)(3.1,1.5)
\psline(2.25,1)(1.75,1.5)
\psline{->}(2.75,0.5)(2.15,1.1)
\psline(1.6,0.5)(2.75,0.5)
\psline{->}(0.5,0.5)(1.7,0.5)
\psdots(2.75,0.5)(1.75,1.5)
\psarc[linestyle=dashed](2.75,1.5){1}{-180}{-90}
\psline[linestyle=dotted](2.75,0)(2.75,2)
\endpspicture \\[9mm]
&=& \pspicture[0.4](-0.1,0)(4.6,1.5)
\psline(3.5,0.5)(4,0.5)
\psline{->}(2.25,0.5)(3.6,0.5)
\psline{->}(1,0.5)(2.35,0.5)
\psline{->}(0.5,0.5)(1.1,0.5)
\psdots(1.5,0.5)(3,0.5)
\psarc[linestyle=dashed](2.25,0.2){0.8}{20}{160}
\psline[linestyle=dotted](3,0)(3,1.5)
\endpspicture + \pspicture[0.4](-0.1,0)(4.6,2)
\psline(3,1.5)(4,1.5)
\psline{->}(1.75,1.5)(3.1,1.5)
\psline(2.25,1)(1.75,1.5)
\psline{->}(2.75,0.5)(2.15,1.1)
\psline(1.6,0.5)(2.75,0.5)
\psline{->}(0.5,0.5)(1.7,0.5)
\psdots(2.75,0.5)(1.75,1.5)
\psarc[linestyle=dashed](2.75,1.5){1}{-180}{-90}
\psline[linestyle=dotted](1.75,0)(1.75,2)
\endpspicture \:\:=\:\: 
\pspicture[0.4](-0.1,0)(4.6,1.5)
\psline(3.5,0.5)(4,0.5)
\psline{->}(2.25,0.5)(3.6,0.5)
\psline{->}(1,0.5)(2.35,0.5)
\psline{->}(0.5,0.5)(1.1,0.5)
\psdots(1.5,0.5)(3,0.5)
\psline[linestyle=dotted](3,0.5)(3,1.5)
\psarc[linestyle=dashed](2.25,0.2){0.8}{20}{160}
\endpspicture
\end{eqnarray*}
\end{center}
\caption{The translation in time for the linked diagrams of Fig.\ 
\ref{onepfig} as a consequence of energy conservation implied by momentum
conservation in this case. In the last Feynman diagram, the vertex to the 
right is still fixed at $t = 0$, while the position of the left vertex is 
to be integrated over $t$ from $- \infty$ to $\infty$. \label{shiftfig}}
\end{figure}

The integration over $\bf x$ and ${\bf x}'$ enforces 3--momentum conservation
also for the linked diagrams in Fig.\ \ref{onepfig} (the third and fourth
contribution in Eq.\ (\ref{onepcor})), so that these contributions are
diagonal in the basis $\{ | {\bf p}_A \rangle \}$ as well. 
Momentum conservation has another important
consequence in the special case of one--particle states: it implies energy
conservation, i.e., $\omega_{{\bf p}_A}^A = \omega_{{\bf p}_A'}^A$. The
product of the wave functions in Eq.\ (\ref{onepcor}) and consequently the
complete matrix elements then become invariant under a translation in time
$t$. One can use this fact to move the right vertex in the last diagram in 
Fig.\ \ref{onepfig} to time $-t$, keeping the left vertex fixed at $t = 0$, 
and as a result the two linked contributions now sum up to a proper (not 
time--ordered) Feynman diagram. The whole process is represented 
diagrammatically in Fig.\ \ref{shiftfig}.

In mathematical terms, we replace $t$ by $-t$, 
interchange $\bf x$ and ${\bf x}'$ and make use of $\omega_{{\bf p}_A}^A = 
\omega_{{\bf p}_A'}^A$, in order to rewrite the contribution corresponding
to the last diagram in Fig.\ \ref{onepfig} as
\be
-i g^2 \int_0^\infty dt \, e^{- \epsilon |t|}
\int d^3 x \, d^3 x' \psi_{{\bf p}_A}^{A \, \ast} (0, {\bf x}) 
\Delta_F^A (0 - t, {\bf x} - {\bf x}') \Delta_F (0 - t, {\bf x} - {\bf x}')
\psi_{{\bf p}_A'}^A (t, {\bf x}') \:.
\ee
It is hence of the same form as the contribution of the other linked 
diagram, except for the range of integration of $t$. Adding up these two
contributions, with the use of the covariant representation Eq.\ 
(\ref{feynmcov}) of the Feynman propagators and
\be
\int_{-\infty}^{\infty} dt \, e^{- \epsilon |t|} e^{i \omega t}
= \frac{i}{\omega + i \epsilon} - \frac{i}{\omega - i \epsilon}
= 2 \pi \delta (\omega)
\ee
in the limit $\epsilon \to 0$, leads to the result
\be
\frac{1}{2 \omega_{{\bf p}_A}^A} \, G^{(2)} (\omega_{{\bf p}_A}^A, 
{\bf p}_A) \, (2 \pi)^3 \delta ({\bf p}_A - {\bf p}_A') \:, 
\label{masscor}
\ee
where
\be
G^{(2)} (p) = i g^2 \int \frac{d^4 k}{(2 \pi)^4} \, \frac{1}{k^2 - \mu^2 + i
\epsilon} \, \frac{1}{(p - k)^2 - m_A^2 + i \epsilon}
\ee
is the second--order contribution to the usual two--point function (without
the $\delta$--function for energy and momentum conservation).
It is very important that we can identify the sum of the linked
diagrams with a proper Feynman diagram as it appears in covariant perturbation
theory, because it is then possible to apply the usual renormalization
procedure in order to make sense of the (without regularization formally
infinite) mathematical expressions. 

As is well--known, $G^{(2)} (p)$ actually only depends on the invariant square
$p^2$, hence on $m_A^2$ in Eq.\ (\ref{masscor}). Putting
\be
\Delta m^2_A = G^{(2)} (m_A^2) \:, \label{massren}
\ee
$\Delta m^2_A$ is of second order in $g$, and the complete matrix element of
$H_{BW}$ up to second order in the one--particle sector reads
\be
\langle {\bf p}_A | H_{BW} | {\bf p}_A' \rangle &=& \left( E_\Omega +
\omega_{{\bf p}_A}^A + \frac{\Delta m^2_A}{2 \omega_{{\bf p}_A}^A} \right)
(2 \pi)^3 \delta ({\bf p}_A - {\bf p}_A') + {\cal O} (g^4) \nn \\
&=& \left( E_\Omega + \sqrt{M_A^2 + {\bf p}_A^2} \right)
(2 \pi)^3 \delta ({\bf p}_A - {\bf p}_A') + {\cal O} (g^4) \:,
\ee
where we have defined the ``renormalized mass'' $M_A$ by
\be
M_A^2 = m_A^2 + \Delta m_A^2 \:. \label{massdef}
\ee
As is clear from the definition Eq.\ (\ref{massren})
of $\Delta m_A^2$, this renormalization corresponds to an on--shell scheme.
To sum up the discussion of the one--particle sector, the second--order
contribution to the Bloch--Wilson Hamiltonian contains the correction to the 
vacuum energy and the mass renormalization to this order. It is clear that
exactly the same conclusions hold for the sector with one $B$--particle.

\begin{figure}
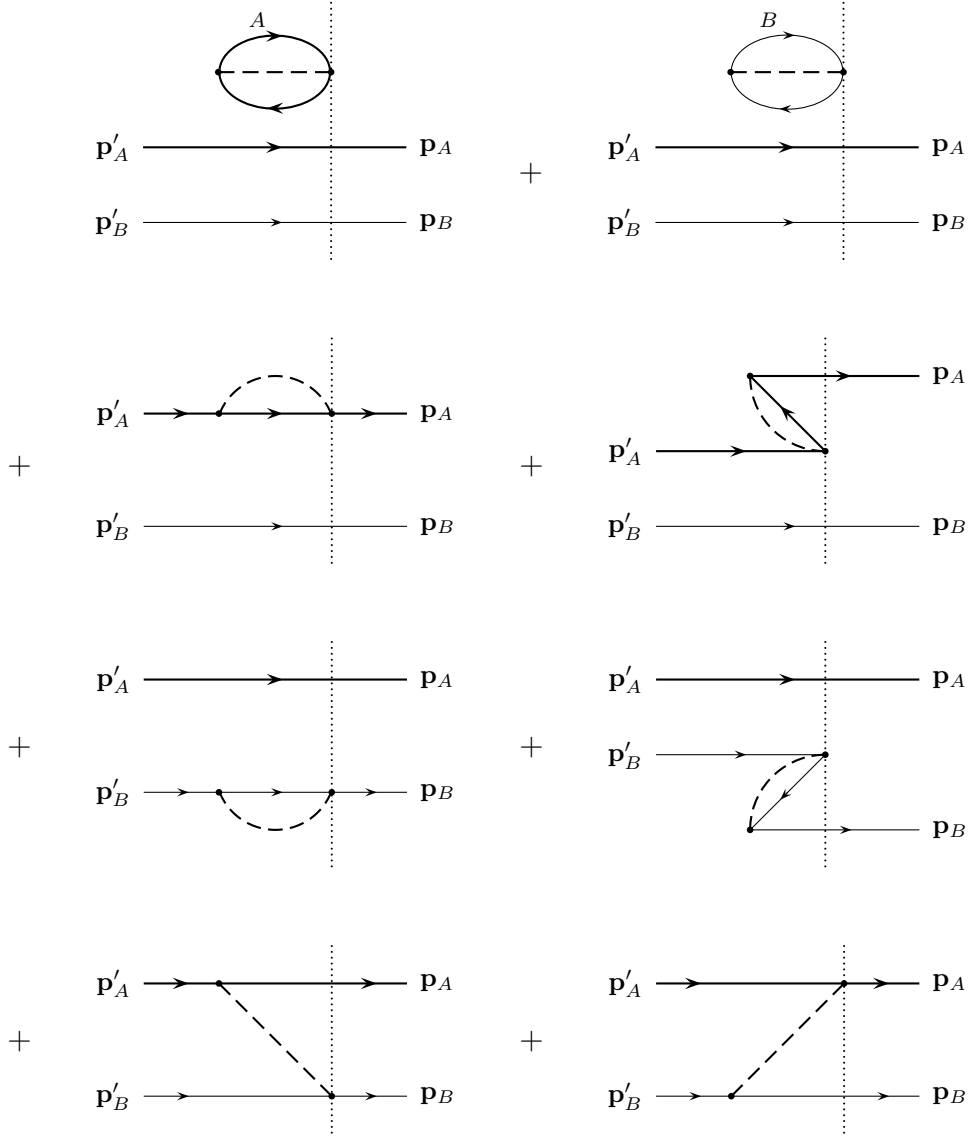

\begin{center} 
\begin{eqnarray*}
& & \pspicture[0.3](-1.4,0)(5.4,3.5)
\psellipse(2.25,2.5)(0.75,0.5)
\psline{->}(2.25,2.98)(2.35,2.98)
\uput[135](2.25,2.98){\mbox{\scriptsize $A$}}
\psline{->}(2.25,2.02)(2.15,2.02)
\psdots(1.5,2.5)(3,2.5)
\psline[linestyle=dashed](1.5,2.5)(3,2.5)
\psline(2.25,1.5)(4,1.5)
\uput[0](4,1.5){\mbox{\footnotesize ${\bf p}_A$}}
\psline{->}(0.5,1.5)(2.35,1.5)
\uput[180](0.5,1.5){\mbox{\footnotesize ${\bf p}_A'$}}
\psline[linewidth=.4pt](2.25,0.5)(4,0.5)
\uput[0](4,0.5){\mbox{\footnotesize ${\bf p}_B$}}
\psline[linewidth=.4pt]{->}(0.5,0.5)(2.35,0.5)
\uput[180](0.5,0.5){\mbox{\footnotesize ${\bf p}_B'$}}
\psline[linestyle=dotted](3,0)(3,3.5)
\endpspicture + \pspicture[0.3](-0.9,0)(5.4,3.5)
\psellipse[linewidth=.4pt](2.25,2.5)(0.75,0.5)
\psline[linewidth=.4pt]{->}(2.25,2.99)(2.35,2.99)
\uput[135](2.25,2.98){\mbox{\scriptsize $B$}}
\psline[linewidth=.4pt]{->}(2.25,2.01)(2.15,2.01)
\psdots(1.5,2.5)(3,2.5)
\psline[linestyle=dashed](1.5,2.5)(3,2.5)
\psline(2.25,1.5)(4,1.5)
\uput[0](4,1.5){\mbox{\footnotesize ${\bf p}_A$}}
\psline{->}(0.5,1.5)(2.35,1.5)
\uput[180](0.5,1.5){\mbox{\footnotesize ${\bf p}_A'$}}
\psline[linewidth=.4pt](2.25,0.5)(4,0.5)
\uput[0](4,0.5){\mbox{\footnotesize ${\bf p}_B$}}
\psline[linewidth=.4pt]{->}(0.5,0.5)(2.35,0.5)
\uput[180](0.5,0.5){\mbox{\footnotesize ${\bf p}_B'$}}
\psline[linestyle=dotted](3,0)(3,3.5)
\endpspicture
\\[9mm]
& & {} + \pspicture[0.4](-0.9,0)(5.4,3)
\psline(3.5,2)(4,2)
\uput[0](4,2){\mbox{\footnotesize ${\bf p}_A$}}
\psline{->}(2.25,2)(3.6,2)
\psline{->}(1,2)(2.35,2)
\psline{->}(0.5,2)(1.1,2)
\uput[180](0.5,2){\mbox{\footnotesize ${\bf p}_A'$}}
\psdots(1.5,2)(3,2)
\psarc[linestyle=dashed](2.25,1.7){0.8}{20}{160}
\psline[linewidth=.4pt](2.25,0.5)(4,0.5)
\uput[0](4,0.5){\mbox{\footnotesize ${\bf p}_B$}}
\psline[linewidth=.4pt]{->}(0.5,0.5)(2.35,0.5)
\uput[180](0.5,0.5){\mbox{\footnotesize ${\bf p}_B'$}}
\psline[linestyle=dotted](3,0)(3,3)
\endpspicture + \pspicture[0.4](-0.9,0)(5.4,3)
\psline(3,2.5)(4,2.5)
\uput[0](4,2.5){\mbox{\footnotesize ${\bf p}_A$}}
\psline{->}(1.75,2.5)(3.1,2.5)
\psline(2.25,2)(1.75,2.5)
\psline{->}(2.75,1.5)(2.15,2.1)
\psline(1.6,1.5)(2.75,1.5)
\psline{->}(0.5,1.5)(1.7,1.5)
\uput[180](0.5,1.5){\mbox{\footnotesize ${\bf p}_A'$}}
\psdots(2.75,1.5)(1.75,2.5)
\psarc[linestyle=dashed](2.75,2.5){1}{-180}{-90}
\psline[linewidth=.4pt](2.25,0.5)(4,0.5)
\uput[0](4,0.5){\mbox{\footnotesize ${\bf p}_B$}}
\psline[linewidth=.4pt]{->}(0.5,0.5)(2.35,0.5)
\uput[180](0.5,0.5){\mbox{\footnotesize ${\bf p}_B'$}}
\psline[linestyle=dotted](2.75,0)(2.75,3)
\endpspicture
\\[9mm]
& & {} + \pspicture[0.5](-0.9,0)(5.4,3)
\psline(2.25,2.5)(4,2.5)
\uput[0](4,2.5){\mbox{\footnotesize ${\bf p}_A$}}
\psline{->}(0.5,2.5)(2.35,2.5)
\uput[180](0.5,2.5){\mbox{\footnotesize ${\bf p}_A'$}}
\psline[linewidth=.4pt](3.5,1)(4,1)
\uput[0](4,1){\mbox{\footnotesize ${\bf p}_B$}}
\psline[linewidth=.4pt]{->}(2.25,1)(3.6,1)
\psline[linewidth=.4pt]{->}(1,1)(2.35,1)
\psline[linewidth=.4pt]{->}(0.5,1)(1.1,1)
\uput[180](0.5,1){\mbox{\footnotesize ${\bf p}_B'$}}
\psdots(1.5,1)(3,1)
\psarc[linestyle=dashed](2.25,1.3){0.8}{-160}{-20}
\psline[linestyle=dotted](3,0)(3,3)
\endpspicture + \pspicture[0.5](-0.9,0)(5.4,3)
\psline(2.25,2.5)(4,2.5)
\uput[0](4,2.5){\mbox{\footnotesize ${\bf p}_A$}}
\psline{->}(0.5,2.5)(2.35,2.5)
\uput[180](0.5,2.5){\mbox{\footnotesize ${\bf p}_A'$}}
\psline[linewidth=.4pt](3,0.5)(4,0.5)
\uput[0](4,0.5){\mbox{\footnotesize ${\bf p}_B$}}
\psline[linewidth=.4pt]{->}(1.75,0.5)(3.1,0.5)
\psline[linewidth=.4pt](2.25,1)(1.75,0.5)
\psline[linewidth=.4pt]{->}(2.75,1.5)(2.15,0.9)
\psline[linewidth=.4pt](1.6,1.5)(2.75,1.5)
\psline[linewidth=.4pt]{->}(0.5,1.5)(1.7,1.5)
\uput[180](0.5,1.5){\mbox{\footnotesize ${\bf p}_B'$}}
\psdots(2.75,1.5)(1.75,0.5)
\psarc[linestyle=dashed](2.75,0.5){1}{90}{180}
\psline[linestyle=dotted](2.75,0)(2.75,3)
\endpspicture
\\[9mm]
& & {} + \pspicture[0.45](-0.9,0)(5.4,2.5)
\psline(3.5,2)(4,2)
\uput[0](4,2){\mbox{\footnotesize ${\bf p}_A$}}
\psline{->}(1,2)(3.6,2)
\psline{->}(0.5,2)(1.1,2)
\uput[180](0.5,2){\mbox{\footnotesize ${\bf p}_A'$}}
\psline[linewidth=.4pt](3.5,0.5)(4,0.5)
\uput[0](4,0.5){\mbox{\footnotesize ${\bf p}_B$}}
\psline[linewidth=.4pt]{->}(1,0.5)(3.6,0.5)
\psline[linewidth=.4pt]{->}(0.5,0.5)(1.1,0.5)
\uput[180](0.5,0.5){\mbox{\footnotesize ${\bf p}_B'$}}
\psdots(1.5,2)(3,0.5)
\psline[linestyle=dashed](1.5,2)(3,0.5)
\psline[linestyle=dotted](3,0)(3,2.5)
\endpspicture + \pspicture[0.45](-0.9,0)(5.4,2.5)
\psline(3.5,2)(4,2)
\uput[0](4,2){\mbox{\footnotesize ${\bf p}_A$}}
\psline{->}(1,2)(3.6,2)
\psline{->}(0.5,2)(1.1,2)
\uput[180](0.5,2){\mbox{\footnotesize ${\bf p}_A'$}}
\psline[linewidth=.4pt](3.5,0.5)(4,0.5)
\uput[0](4,0.5){\mbox{\footnotesize ${\bf p}_B$}}
\psline[linewidth=.4pt]{->}(1,0.5)(3.6,0.5)
\psline[linewidth=.4pt]{->}(0.5,0.5)(1.1,0.5)
\uput[180](0.5,0.5){\mbox{\footnotesize ${\bf p}_B'$}}
\psdots(1.5,0.5)(3,2)
\psline[linestyle=dashed](1.5,0.5)(3,2)
\psline[linestyle=dotted](3,0)(3,2.5)
\endpspicture
\end{eqnarray*}
\end{center}
\caption{The second--order contributions in the two--particle sector.
The first six diagrams are essentially those of Figs.\ \ref{vacfig} and
\ref{onepfig}, except that they are multiplied with further propagators
which merely introduce additional $\delta$--functions in the external
momenta. The last two diagrams describe the interaction of the two particles
to this order and correspond to the expression Eq.\ (\ref{twopcor}). 
\label{twopfig}}
\end{figure}

We will now return to the two--particle subspace $\Omega_0$ defined in
Eq.\ (\ref{omtwop}). The diagrams corresponding to the matrix elements of the
Bloch--Wilson Hamiltonian in this subspace are presented in Fig.\ 
\ref{twopfig}. 
They can be naturally classified into unlinked diagrams (the 
first two), linked but disconnected diagrams (the following four diagrams)
and connected diagrams (the last two).

It should be clear by now that the first six contributions
are diagonal in the basis $\{ | {\bf p}_A, {\bf p}_B \rangle \}$, and that,
together with the zero--order contribution, they lead to the result
\be
\left( E_\Omega + \sqrt{M_A^2 + {\bf p}_A^2} + \sqrt{M_B^2 + {\bf p}_B^2}
\right) (2 \pi)^3 \delta ({\bf p}_A - {\bf p}_A') (2 \pi)^3 \delta 
({\bf p}_B - {\bf p}_B') \:,
\ee
with $E_\Omega$ and $M_A^2$ (and analogously $M_B^2$) as defined in 
Eqs.\ (\ref{vaccorr}) and (\ref{massdef}), respectively. In general, unlinked
diagrams contribute to the vacuum energy, while linked but disconnected
diagrams are concerned with the properties of the individual particles (in
the case of two--particle states).

As for the last two diagrams in Fig.\ \ref{twopfig}, they correspond to the
mathematical expression
\be
\lefteqn{\langle {\bf p}_A, {\bf p}_B | V | {\bf p}_A', {\bf p}_B' \rangle}
\nn \\[2mm]
&=& -i g^2 \int_{-\infty}^0 dt \, e^{- \epsilon |t|}
\int d^3 x \, d^3 x' \psi_{{\bf p}_B}^{B \, \ast} (0, {\bf x}) 
\psi_{{\bf p}_B'}^B (0, {\bf x}) \Delta_F (0 - t, {\bf x} - {\bf x}')
\psi_{{\bf p}_A}^{A \, \ast} (t, {\bf x}') \psi_{{\bf p}_A'}^A (t, {\bf x}') 
\nn \\
& & {}- i g^2 \int_{-\infty}^0 dt \, e^{- \epsilon |t|}
\int d^3 x \, d^3 x' \psi_{{\bf p}_A}^{A \, \ast} (0, {\bf x}) 
\psi_{{\bf p}_A'}^A (0, {\bf x}) \Delta_F (0 - t, {\bf x} - {\bf x}')
\psi_{{\bf p}_B}^{B \, \ast} (t, {\bf x}') \psi_{{\bf p}_B'}^B (t, {\bf x}') 
\nn \\
& & \label{twopcor}
\ee
with the wave functions $\psi_{\bf p}^A (t, {\bf x})$ defined as in 
Eq.\ (\ref{wavef}) (and analogously for the $B$--particles). The integrations
over the vertex positions $\bf x$ and ${\bf x}'$ enforce conservation of
the total momentum ${\bf p}_A + {\bf p}_B = {\bf p}_A' + {\bf p}_B'$. Of
course, in this case there is no conservation of the individual momenta,
and so ``energy conservation'' is {\em not}\/ implied, i.e., 
$\omega_{{\bf p}_A}^A + \omega_{{\bf p}_B}^B \neq \omega_{{\bf p}_A'}^A
+ \omega_{{\bf p}_B'}^B$ in general. Physically, this means that states with
different unperturbed energies mix, which was obviously to be expected in the 
case of bound states. As a consequence, Eq.\ (\ref{twopcor}) is not 
invariant under time translations, and there is no way to combine the 
corresponding diagrams into a proper on--shell Feynman diagram.

We conclude that it is the connected diagrams that determine the
interaction of the constituents of a bound state 
(in the case of two--particle 
states). In principle, it was necessary to determine the energies of the
vacuum and the one--particle states to be able to identify that part of the
energy of the two--particle states which has to do with the binding (or with
the kinetic energy of the particles far away from the scattering region in
the case of scattering states). However, the diagrammatic analysis allows
to identify immediately the contributions which are related to the
interaction of the constituents.

We can now write down the Schr\"odinger equation corresponding to the
Hamiltonian $H_{BW}$ in the two--particle sector,
\be
\lefteqn{\left( \sqrt{M_A^2 + {\bf p}_A^2} + \sqrt{M_B^2 + {\bf p}_B^2} 
\right) \psi ({\bf p}_A, {\bf p}_B)} \nn \\
& & {}+ \int \frac{d^3 p_A'}{(2 \pi)^3}
\frac{d^3 p_B'}{(2 \pi)^3} \langle {\bf p}_A, {\bf p}_B | V 
| {\bf p}_A', {\bf p}_B' \rangle \psi ({\bf p}_A', {\bf p}_B') =
\left( E - E_\Omega \right) \psi ({\bf p}_A, {\bf p}_B) \:,
\label{twopschroed}
\ee
with the momentum--space wave function
\be
\psi ({\bf p}_A, {\bf p}_B) = \langle {\bf p}_A, {\bf p}_B | \psi \rangle \:.
\ee
By use of the non--covariant form Eq.\ (\ref{feynmnon}) of the Feynman 
propagators, the potential Eq.\ (\ref{twopcor}) can be cast in the form
\be
\lefteqn{\langle {\bf p}_A, {\bf p}_B | V | {\bf p}_A', {\bf p}_B' \rangle}
\nn \\[2mm]
&=& - \frac{g^2}{\sqrt{2 \omega_{{\bf p}_A}^A 2 \omega_{{\bf p}_B}^B
2 \omega_{{\bf p}_A'}^A 2 \omega_{{\bf p}_B'}^B}} \times \nn \\
& & \frac{1}
{2 \omega_{{\bf p}_A - {\bf p}_A'}} \left( \frac{1}{\omega_{{\bf p}_A}^A
+ \omega_{{\bf p}_A - {\bf p}_A'} - \omega_{{\bf p}_A'}^A - i \epsilon}
+ \frac{1}{\omega_{{\bf p}_B}^B + \omega_{{\bf p}_B - {\bf p}_B'} - 
\omega_{{\bf p}_B'}^B - i \epsilon} \right) \nn \\[2mm]
& & {}\times (2 \pi)^3 \delta ({\bf p}_A
+ {\bf p}_B - {\bf p}_A' - {\bf p}_B') \:. \label{potential}
\ee
In the latter expression, $\epsilon$ can be put to zero, since for 
$\mu \neq 0$ the denominators are strictly positive, and for $\mu = 0$
the singularity is integrable and does not lead to any imaginary contribution.
The singularity for $\mu = 0$ is of a similar nature as the 
one appearing in the usual Coulomb potential in momentum space \cite{BS57}.  
It is associated with a long--range force due to a massless exchange particle. 
In the spirit of perturbative renormalization, it is permissible 
to replace the ``bare'' masses $m_A$ and $m_B$ in the potential 
Eq.\ (\ref{potential}) by their renormalized counterparts, given that the 
difference between the two is of higher order in the coupling constant. 

One can now make use of the overall momentum conservation in order to reduce
the dynamics to the center--of--mass system. One then has, with ${\bf p} =
{\bf p}_A = - {\bf p}_B$,
\be
\left( \sqrt{M_A^2 + {\bf p}^2} + \sqrt{M_B^2 + {\bf p}^2} \right) 
\psi ({\bf p}) + \int \frac{d^3 p'}{(2 \pi)^3} \, V({\bf p}, {\bf p}')
\psi ({\bf p}') = E' \psi ({\bf p}) \:, \label{schroedinger}
\ee
where
\be
E' = E - E_\Omega
\ee
and
\be
V({\bf p}, {\bf p}') = - \frac{g^2}{\sqrt{2 \omega_{{\bf p}}^A 
2 \omega_{{\bf p}}^B 2 \omega_{{\bf p}'}^A 2 \omega_{{\bf p}'}^B}} \, 
\frac{1}{2 \omega_{{\bf p} - {\bf p}'}} \left( \frac{1}{\omega_{{\bf p}}^A
+ \omega_{{\bf p} - {\bf p}'} - \omega_{{\bf p}'}^A}
+ \frac{1}{\omega_{{\bf p}}^B + \omega_{{\bf p} - {\bf p}'} - 
\omega_{{\bf p}'}^B} \right) \:. \hspace{5mm} \label{relpot}
\ee
In this form, the equation will be solved in Section \ref{sec:num}. The 
effective Hamiltonian on the left--hand side of the Schr\"odinger equation
Eq.\ (\ref{schroedinger}) obviously consists of a relativistic kinetic term 
and a potential term, where the operator corresponding to the potential is
{\em non--local and non--Hermitian}. The non--relativistic and one--body
limits of the potential will be discussed in the following section.

\begin{figure}[t]
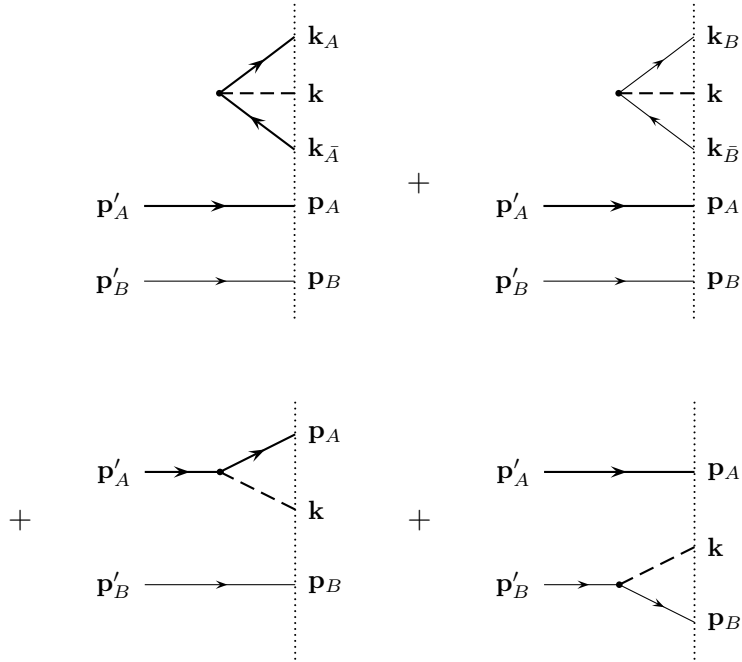

\begin{center} 
\begin{eqnarray*}
& & \pspicture[0.4](-1.4,0)(3.9,4.25)
\psline(2,3.37)(2.5,3.75)
\uput[0](2.5,3.75){\mbox{\footnotesize ${\bf k}_A$}}
\psline{->}(1.5,3)(2.1,3.45)
\psline(2,2.62)(1.5,3)
\psline{->}(2.5,2.25)(1.9,2.7)
\uput[0](2.5,2.25){\mbox{\footnotesize ${\bf k}_{\bar{A}}$}}
\psline[linestyle=dashed](1.5,3)(2.5,3)
\uput[0](2.5,3){\mbox{\footnotesize ${\bf k}$}}
\psdots(1.5,3)
\psline(1.5,1.5)(2.5,1.5)
\uput[0](2.5,1.5){\mbox{\footnotesize ${\bf p}_A$}}
\psline{->}(0.5,1.5)(1.6,1.5)
\uput[180](0.5,1.5){\mbox{\footnotesize ${\bf p}_A'$}}
\psline[linewidth=.4pt](1.5,0.5)(2.5,0.5)
\uput[0](2.5,0.5){\mbox{\footnotesize ${\bf p}_B$}}
\psline[linewidth=.4pt]{->}(0.5,0.5)(1.6,0.5)
\uput[180](0.5,0.5){\mbox{\footnotesize ${\bf p}_B'$}}
\psline[linestyle=dotted](2.5,0)(2.5,4.25)
\endpspicture + \pspicture[0.4](-0.9,0)(3.9,4.25)
\psline[linewidth=.4pt](2,3.37)(2.5,3.75)
\uput[0](2.5,3.75){\mbox{\footnotesize ${\bf k}_B$}}
\psline[linewidth=.4pt]{->}(1.5,3)(2.1,3.45)
\psline[linewidth=.4pt](2,2.62)(1.5,3)
\psline[linewidth=.4pt]{->}(2.5,2.25)(1.9,2.7)
\uput[0](2.5,2.25){\mbox{\footnotesize ${\bf k}_{\bar{B}}$}}
\psline[linestyle=dashed](1.5,3)(2.5,3)
\uput[0](2.5,3){\mbox{\footnotesize ${\bf k}$}}
\psdots(1.5,3)
\psline(1.5,1.5)(2.5,1.5)
\uput[0](2.5,1.5){\mbox{\footnotesize ${\bf p}_A$}}
\psline{->}(0.5,1.5)(1.6,1.5)
\uput[180](0.5,1.5){\mbox{\footnotesize ${\bf p}_A'$}}
\psline[linewidth=.4pt](1.5,0.5)(2.5,0.5)
\uput[0](2.5,0.5){\mbox{\footnotesize ${\bf p}_B$}}
\psline[linewidth=.4pt]{->}(0.5,0.5)(1.6,0.5)
\uput[180](0.5,0.5){\mbox{\footnotesize ${\bf p}_B'$}}
\psline[linestyle=dotted](2.5,0)(2.5,4.25)
\endpspicture
\\[9mm]
& & {} + \pspicture[0.5](-0.9,-0.5)(3.9,3)
\psline(2,2.25)(2.5,2.5)
\uput[0](2.5,2.5){\mbox{\footnotesize ${\bf p}_A$}}
\psline{->}(1.5,2)(2.1,2.3)
\psline(1,2)(1.55,2)
\psline{->}(0.5,2)(1.1,2)
\uput[180](0.5,2){\mbox{\footnotesize ${\bf p}_A'$}}
\psdots(1.5,2)
\psline[linestyle=dashed](1.5,2)(2.5,1.5)
\uput[0](2.5,1.5){\mbox{\footnotesize ${\bf k}$}}
\psline[linewidth=.4pt](1.5,0.5)(2.5,0.5)
\uput[0](2.5,0.5){\mbox{\footnotesize ${\bf p}_B$}}
\psline[linewidth=.4pt]{->}(0.5,0.5)(1.6,0.5)
\uput[180](0.5,0.5){\mbox{\footnotesize ${\bf p}_B'$}}
\psline[linestyle=dotted](2.5,-0.5)(2.5,3)
\endpspicture + \pspicture[0.5](-0.9,-0.5)(3.9,3)
\psline(1.5,2)(2.5,2)
\uput[0](2.5,2){\mbox{\footnotesize ${\bf p}_A$}}
\psline{->}(0.5,2)(1.6,2)
\uput[180](0.5,2){\mbox{\footnotesize ${\bf p}_A'$}}
\psdots(1.5,0.5)
\psline[linestyle=dashed](1.5,0.5)(2.5,1)
\uput[0](2.5,1){\mbox{\footnotesize $\bf k$}}
\psline[linewidth=.4pt](2,0.25)(2.5,0)
\uput[0](2.5,0){\mbox{\footnotesize ${\bf p}_B$}}
\psline[linewidth=.4pt]{->}(1.5,0.5)(2.1,0.2)
\psline[linewidth=.4pt](1,0.5)(1.5,0.5)
\psline[linewidth=.4pt]{->}(0.5,0.5)(1.1,0.5)
\uput[180](0.5,0.5){\mbox{\footnotesize ${\bf p}_B'$}}
\psline[linestyle=dotted](2.5,-0.5)(2.5,3)
\endpspicture
\end{eqnarray*}
\end{center}
\caption{The first--order contributions to the physical states 
in the two--particle sector.
The correspondence to mathematical expressions is as before, except that
the ingoing external lines have to be contracted with the solutions 
$\psi ({\bf p}_A, {\bf p}_B)$ of the Schr\"odinger equation 
Eq.\ (\ref{twopschroed}), and the particle content of the state can be read off
from the outgoing external lines.
\label{statefig}}
\end{figure}

Finally, we will turn our attention to the physical {\em states}\/ in the
perturbative expansion. They are simply given by the application of
\be
U_{BW} = {\bf 1} - i \int_{-\infty}^0 dt \, e^{- \epsilon |t|}
H_I (t) + {\cal O} (H_I^2) \label{ubwfirst}
\ee 
to the two--particle states
\be
| \psi \rangle = \int \frac{d^3 p_A}{(2 \pi)^3} \frac{d^3 p_B}{(2 \pi)^3}
\, \psi ({\bf p}_A, {\bf p}_B) | {\bf p}_A, {\bf p}_B \rangle \in \Omega_0 \:,
\ee
with the wave functions $\psi ({\bf p}_A, {\bf p}_B)$ which solve
the Schr\"odinger equation Eq.\ (\ref{twopschroed}). The result is easily 
written down and is represented in Fig.\ \ref{statefig}.
As an example, the mathematical expression corresponding to the last diagram
in Fig.\ \ref{statefig} is
\be
& & - i g \int \frac{d^3 p_A'}{(2 \pi)^3}
\frac{d^3 p_B'}{(2 \pi)^3} \, \psi ({\bf p}_A', {\bf p}_B')
\int_{-\infty}^0 dt \, e^{-\epsilon |t|} \int d^3 x \nn \\
& & \times \int \frac{d^3 p_A}{(2 \pi)^3} \frac{d^3 p_B}{(2 \pi)^3}
\frac{d^3 k}{(2 \pi)^3} \, \psi^{B \, \ast}_{{\bf p}_B} (t, {\bf x})
\psi^\ast_{\bf k} (t, {\bf x}) \psi^B_{{\bf p}_B'} (t, {\bf x})
(2 \pi)^3 \delta ({\bf p}_A - {\bf p}_A') | {\bf p}_A, {\bf p}_B, {\bf k}
\rangle \nn \\[2mm]
&=& - g \int \frac{d^3 p_A'}{(2 \pi)^3}
\frac{d^3 p_B'}{(2 \pi)^3} \, \psi ({\bf p}_A', {\bf p}_B') \nn \\ 
& & {}\times \frac{1}{\sqrt{2 \omega^B_{{\bf p}_B'}}} 
\int \frac{d^3 k}{(2 \pi)^3}
\frac{1}{\sqrt{2 \omega^B_{{\bf p}_B' - {\bf k}} 2 \omega_{\bf k}}}
\, \frac{1}{\omega^B_{{\bf p}_B' - {\bf k}} + \omega_{\bf k} - 
\omega^B_{{\bf p}_B'}} \, | {\bf p}_A', {\bf p}_B' - {\bf k}, {\bf k} 
\rangle \:.
\ee
For the model considered, the operator $U_{BW}$ is unitary to first order in
the coupling constant. 
As a consequence, the physical states are normalized when
the normalization is calculated to this order, provided that the wave
function $\psi ({\bf p}_A, {\bf p}_B)$ is properly normalized. However, this
is a feature that does not carry over to higher orders, and one should bear
in mind that, on physical grounds, the criterium for a bound state is that the
{\em relative wave function}\/ $\psi({\bf p})$ be normalizable, the latter
describing the probability distribution of the constituents in the
center--of--mass system.

The operator $U_{BW}^\dagger U_{BW}$, which appears naturally when calculating
the norm of the physical states, also plays a r\^ole when trying to make
the effective Hamiltonian hermitian by a similarity transformation. For
future investigations, we only remark here that this operator, when divided
by its vacuum expectation value $\langle \Omega | U_{BW}^\dagger U_{BW}
| \Omega \rangle$, leads to well--defined expressions
even if one includes the second--order term arising from substituting the
expression Eq.\ (\ref{ubwfirst}) for $U_{BW}$. A generalization of the optical
theorem applies to this case, for which the adiabatic $\epsilon$--prescription
proves essential.

\section{Non--Relativistic and One--Body Limits}

In this section, we will carry out an analytic study of the properties of 
the Bloch--Wilson Hamiltonian to lowest non--trivial order, or the 
corresponding Schr\"odinger equation Eq.\ (\ref{schroedinger}). A numerical 
solution of Eq.\ (\ref{schroedinger}) will be presented in the next section, 
which at the same time will confirm the results to be obtained in the 
following.

There are two different limits in which a formalism for bound state 
calculations can easily be tested for consistency: the non--relativistic
limit, in which Eq.\ (\ref{schroedinger}) should reproduce the corresponding
non--relativistic Schr\"odinger equation, and the one--body limit $M_B \to
\infty$, where Eq.\ (\ref{schroedinger}) is expected to reduce to a
(relativistic) equation for particle $A$ in a static potential \cite{Gro82}.

We will discuss the non--relativistic limit first, which is by definition the
case ${\bf p}^2 \ll M_A^2, M_B^2$, where ${\bf p} = {\bf p}_A = -{\bf p}_B$
is the relative momentum. More precisely, this means that there exists a
{\em solution}\/ $\psi ({\bf p})$ of the effective Schr\"odinger equation
which takes on non--negligible values only in this region of momentum space, 
or equivalently, which is strongly suppressed for ${\bf p}^2 \gtrsim M_r^2$,
\be
M_r = \frac{M_A M_B}{M_A + M_B} < M_A, M_B
\ee
being the usual reduced mass. {\em If}\/ such a solution exists, the effective
potential Eq.\ (\ref{relpot}) can be replaced in the Schr\"odinger equation
Eq.\ (\ref{schroedinger}) by its limiting form for
${\bf p}^2 \ll M_A^2, M_B^2$ and ${\bf p}^{\prime \, 2} \ll M_A^2, M_B^2$,
\be
V ({\bf p}, {\bf p}') = - \frac{g^2}{4 M_A M_B} 
\frac{1}{\omega_{{\bf p} - {\bf p}'}^2}
= - \frac{4 \pi \alpha}{\mu^2 + ({\bf p} - {\bf p}')^2} \:, \label{approx}
\ee
with the dimensionless effective coupling constant
\be
\alpha = \frac{g^2}{16 \pi M_A M_B} \:,
\ee
which is the analogue in this model of the fine--structure constant in QED.
To arrive at the approximation Eq.\ (\ref{approx}), we have made use of
\be
\omega_{\bf p}^A + \omega_{{\bf p} - {\bf p}'} - \omega_{{\bf p}'}^A
\approx \frac{{\bf p}^2 - {\bf p}^{\prime \, 2}}{2 M_A} + 
\omega_{{\bf p} - {\bf p}'} \label{approx1}
\ee
and
\be
\frac{|{\bf p}^2 - {\bf p}^{\prime \, 2}|}{2 M_A} \leq
\frac{|{\bf p} - {\bf p}'| |{\bf p} + {\bf p}'|}{2 M_A} \ll 
|{\bf p} - {\bf p}'| \leq \omega_{{\bf p} - {\bf p}'} \:, \label{approx2}
\ee
analogously for $\omega_{\bf p}^B$ and $M_B$.

Consequently, the non--relativistic solutions $\psi ({\bf p})$ of Eq.\ 
(\ref{schroedinger}) are approximate solutions of the corresponding
non--relativistic Schr\"odinger equation
\be
\frac{{\bf p}^2}{2 M_r} \, \psi ({\bf p}) - \int \frac{d^3 p'}{(2 \pi)^3}
\frac{4 \pi \alpha}{\mu^2 + ({\bf p} - {\bf p}')^2} \, \psi ({\bf p}') =
\Big( E' - M_A - M_B \Big) \psi ({\bf p}) \:, \label{yukawa}
\ee
or, written in the more familiar form in position space after a Fourier
transformation,
\be
\left( - \frac{1}{2 M_r} \nabla^2 - \alpha \, \frac{e^{- \mu r}}{r} \right) 
\psi ({\bf r}) = \Big( E' - M_A - M_B \Big) \psi ({\bf r}) \:.
\ee
As is well--known, in the case $\mu = 0$ the ground state solution of Eq.\
(\ref{yukawa}) is
\be
\psi_1 ({\bf p}) = \sqrt{\frac{32 (\alpha M_r)^5}{\pi}} \,
\frac{1}{(\alpha^2 M_r^2 + {\bf p}^2)^2} \:. \label{groundstate}
\ee
For $\alpha \ll 1$, $\psi_1 ({\bf p})$ is strongly suppressed at 
${\bf p}^2 \gtrsim M_r^2$ compared to ${\bf p}^2 \ll M_r^2$, hence 
self--consistency is achieved. The wave functions of the excited states show 
similarly strong ${\bf p}$--dependences. Note, however, that a priori other 
solutions of Eq.\ (\ref{schroedinger}) with weaker ${\bf p}$--dependences are
not excluded for $\alpha \ll 1$, although they are certainly not expected on 
physical grounds. The numerical calculations presented in the next section
therefore play an important r\^ole in the analysis of this limit.

Another way to put the condition of strong ${\bf p}$--dependence is to
refer to the characteristic momentum scale $\alpha M_r$ (from the expectation 
value $\langle |{\bf p}| \rangle$). Considering now the case $\mu \neq 0$,
it follows from phenomenological considerations (extension of the bound state)
that the characteristic momentum is given by the larger of $\alpha M_r$ and 
$\mu$. In particular, for the non--relativistic limit 
one needs that $\mu \ll M_r$, in which case for the wave function a strong 
${\bf p}$--dependence qualitatively similar to Eq.\ (\ref{groundstate}) is 
expected. As a result, the non--relativistic
limit is self--consistent and is realized for $\alpha \ll 1$ and $\mu \ll M_r$,
both conditions being necessary.

The second limit of Eq.\ (\ref{schroedinger}) we will discuss, is the 
so--called one--body limit $M_B \to \infty$. In more physical terms, this
means that $M_B^2 \gg M_A^2$, but also that $M_B^2$ is much larger than 
the characteristic momentum scale $\langle {\bf p}^2 \rangle$, which in
turn is determined by the solutions $\psi ({\bf p})$ of Eq.\ 
(\ref{schroedinger}) in this limit. The situation hence bears some similarity 
with the non--relativistic limit, and in fact one can use the analogues of
Eqs.\ (\ref{approx1}) and (\ref{approx2}) to determine the approximate form of 
the effective potential in this limit. As a result, in this situation Eq.\
(\ref{schroedinger}) can be approximated by
\be
\sqrt{M_A^2 + {\bf p}^2} \: \psi ({\bf p}) + \int \frac{d^3 p'}{(2 \pi)^3} 
\, V_A({\bf p}, {\bf p}') \psi ({\bf p}') = (E' - M_B) \psi ({\bf p}) \:, 
\label{onebody}
\ee 
where
\be
V_A({\bf p}, {\bf p}') = - \frac{g^2}{2 M_B \sqrt{2 \omega_{{\bf p}}^A 
2 \omega_{{\bf p}'}^A}} \, \frac{1}{2 \omega_{{\bf p} - {\bf p}'}} 
\left( \frac{1}{\omega_{{\bf p} - {\bf p}'}}
+ \frac{1}{\omega_{{\bf p}}^A + \omega_{{\bf p} - {\bf p}'} - 
\omega_{{\bf p}'}^A} \right) \:. \label{onebodypot}
\ee
Eq.\ (\ref{onebody}) is a relativistic Schr\"odinger equation for particle $A$
in the (static) potential $V_A$. 

We will now show that Eq.\ (\ref{onebody}) is identical
with the equation for particle $A$ interacting with a fixed source through the
exchange of scalars of mass $\mu$, thus establishing the inner consistency of 
the formalism in the one--body limit. To this end, consider the theory defined
by the Hamiltonian
\be
H' &=& H_0' + H_I' \:, \nn \\[2mm]
H_0' &=& \int d^3 x \left[ \phi_A^\dagger ({\bf x}) (m_A^2 - \nabla^2) 
\phi_A ({\bf x}) + \frac{1}{2} \varphi ({\bf x})
(\mu^2 - \nabla^2) \varphi ({\bf x}) \right] \:, \nn \\[2mm]
H_I' &=& \mbox{\bf :} \int d^3 x \, g \phi_A^\dagger ({\bf x}) 
\phi_A ({\bf x}) \varphi ({\bf x}) \, \mbox{\bf :} \, + \frac{g}{2 M_B} 
\varphi (0) \:.
\ee
The form of $H_I'$ is motivated by considering $\phi_B (x)$ as a classical
(positive--energy) Klein--Gordon field, with probability density
\be
\rho (x) = \phi_B^\ast (x) i \frac{\p}{\p t} \phi_B (x) - 
\phi_B (x) i \frac{\p}{\p t} \phi_B^\ast (x) \:.
\ee
In the case that $M_B^2$ is much larger than the relevant momenta ${\bf p}^2$,
one has approximately
\be
\rho (x) = 2 M_B \phi_B^\ast (x) \phi_B (x) \:.
\ee
A particle localized at ${\bf x} = 0$ thus corresponds to
\be
\phi_B^\ast (x) \phi_B (x) = \frac{1}{2 M_B} \delta ({\bf x}) \:.
\ee
Using the latter formula in the classical counterpart of $H_I$ in Eq.\
(\ref{model}) leads to the interaction Hamiltonian $H_I'$.

Considering the one--particle states as in Eq.\ (\ref{onepproj}), 
the Bloch--Wilson
Hamiltonian corresponding to $H'$ leads, to lowest non--trivial order, to the
diagrams represented in Fig.\ \ref{onebodyfig}.
\begin{figure}[t]
\begin{center} \include{pic6} \end{center}
\caption{The second--order contributions in the one--particle sector
for the interaction with a fixed source, represented by double lines
in the diagrams. Only the last two diagrams lead to an interaction,
given by Eq.\ (\ref{sourcepot}). \label{onebodyfig}}
\end{figure}
The first four diagrams are diagonal in the basis $\{ | {\bf p}_A \rangle \}$.
The first two, unlinked, diagrams reproduce the second--order correction
$E_\Omega' - E_0'$ to the vacuum energy,\footnote{Technically, the static 
source belongs to the vacuum, so that the second diagram in Fig.\ 
\ref{onebodyfig} which appears to be a self--energy correction of the source, 
in fact contributes to the vacuum energy. Diagrammatically, the double line is 
not counted as an external line, consequently the second diagram is considered
unlinked.} while the following linked but disconnected diagrams renormalize 
the mass of the particle to $M_A$ as before. The last two diagrams in Fig.\ 
\ref{onebodyfig} correspond to the interaction of the particle with the 
source, and translate to the expression
\be
\langle {\bf p}_A | V_A | {\bf p}_A' \rangle
&=& -i \frac{g^2}{2 M_B} \int_{-\infty}^0 dt \, e^{- \epsilon |t|}
\int d^3 x \, \Delta_F (0 - t, 0 - {\bf x})
\psi_{{\bf p}_A}^{A \, \ast} (t, {\bf x}) \psi_{{\bf p}_A'}^A (t, {\bf x}) 
\nn \\
& & {}- i \frac{g^2}{2 M_B} \int_{-\infty}^0 dt \, e^{- \epsilon |t|}
\int d^3 x \, \psi_{{\bf p}_A}^{A \, \ast} (0, {\bf x}) 
\psi_{{\bf p}_A'}^A (0, {\bf x}) \Delta_F (0 - t, {\bf x} - 0) \:. 
\hspace{5mm} \label{sourcepot}
\ee
It is reassuring to see that this potential can be obtained by substituting
\be
\psi^{B \, \ast}_{{\bf p}_B} (t, {\bf x}) \psi^B_{{\bf p}_B'} (t, {\bf x}) 
= \frac{1}{2 M_B} \delta ({\bf x})
\ee
for the wave function of the $B$--particle in the potential Eq.\ 
(\ref{twopcor}) previously obtained for the two--particle sector.

The corresponding Schr\"odinger equation is
\be
\sqrt{M_A^2 + {\bf p}^2} \: \psi ({\bf p}) + \int \frac{d^3 p'}{(2 \pi)^3} 
\langle {\bf p} | V_A | {\bf p}' \rangle \psi ({\bf p}') = 
(E - E_\Omega') \psi ({\bf p}) \:. \label{sourceschroed}
\ee
Using the non--covariant form Eq.\ (\ref{feynmnon}) of the Feynman 
propagators, it is readily shown that
\be 
\langle {\bf p} | V_A | {\bf p}' \rangle = V_A ({\bf p}, {\bf p}') \:,
\ee
the potential Eq.\ (\ref{onebodypot}) obtained in the one--body limit of Eq.\ 
(\ref{schroedinger}). Except for a constant shift in the energies, Eqs.\
(\ref{onebody}) and (\ref{sourceschroed}) are hence identical, thus
establishing the consistency of the formalism in the one--body limit.

\section{Solutions \label{sec:num}}

In this section, we present the numerical solutions of the eigenvalue equation 
Eq.\ (\ref{schroedinger}). We have restricted attention to the case $\mu = 0$
(massless exchange particle), which represents both the physically most 
interesting situation and the most sensible test for the formalism. We also
restricted the calculations to the $s$--wave spectrum.  There is, however,
no problem in considering higher angular--momentum states as well. Due
to the spherical symmetry of the Bloch--Wilson Hamiltonian
different angular--momentum states decouple.
Since the potential depends, in terms of the angular variables, only
on the angle $\theta$ between the vectors ${\bf p}$ and ${\bf p'}$, one
can employ a partial--wave decomposition
\begin{eqnarray}
V(p,p', \cos \theta )& = & \sum_{l=0}^\infty a_l(p,p') P_l(\cos \theta) \nn \\
   & = & \sum_{l,m} \frac{4 \pi}{ 2 l +1} \, a_l(p,p')
     (-1)^m Y_{lm}(\hat{\bf p}) Y^\ast_{lm} (\hat{\bf p}')  \ \ ,
\end{eqnarray}
where $p = |{\bf p}|$, $\hat{\bf p} = {\bf p}/p$, and analogously for 
${\bf p}'$. Although rather cumbersome, analytical expressions for $a_l$ can 
be obtained for arbitrary values of $l$. 
Since the potential is diagonal in the angular momentum quantum numbers
$l$ and $m$, the eigenstates of the Hamiltonian will be eigenstates
of the angular momentum.

For eigenfunctions with zero angular momentum, the analytical integration of 
the angular variables in the potential term in Eq.\ (\ref{schroedinger})
gives
\be
\lefteqn{\int \frac{d^3 p'}{(2 \pi)^3}
\frac{1}{\sqrt{2 \omega_{{\bf p}'}^A 2 \omega_{{\bf p}'}^B}} \,
\frac{1}{2 \omega_{{\bf p} - {\bf p}'}} \, \frac{\psi(|{\bf p}'|)}
{\omega_{{\bf p}}^A + \omega_{{\bf p} - {\bf p}'} - \omega_{{\bf p}'}^A}} 
\nn \\
&=& \frac{1}{4 \pi p} \int_0^\infty \frac{d p'}{2 \pi} \frac{p' \psi(p')}
{\sqrt{2 \omega_{p'}^A 2 \omega_{p'}^B}} \,
\log \left[\frac{\omega^A_p + p + p' - \omega^A_{p'}}
{\omega^A_p + |p-p'| - \omega^A_{p'}} \right] \:,
\ee
The integrand contains a logarithmic singularity, which,
with the proper care, can be integrated easily.
We use an integration method with a fine grid--size around the 
singularity. Furthermore, because of the wide separation of the
relevant scales appearing in the problem in the weak--coupling limit,
we took special care to
distribute the grid points over the full physical range.
In order to determine the spectrum, we calculate
matrix elements of the Hamiltonian with a number of smooth 
basis functions which go into a numerical eigenvalue
routine. In our calculations we have used two different
bases which we specify in the following.

The Coulomb basis is the best basis to analyze
the eigenstates of the Bloch--Wilson Hamiltonian
in the weak--coupling limit. The Coulomb basis consists of
the eigenstates of the non--re\-la\-tiv\-ist\-ic Coulomb 
problem ($n=1,2,3,\ldots$),
\be
\psi_n(p) = \sqrt{\frac{32 n^2}{\pi \kappa^3}} \,
\frac{1}{[(n p/\kappa)^2 +1]^2} \,
C_{n-1}^1\left( \frac{(n p/\kappa)^2 -1}{(n p/\kappa)^2 + 1} \right) \:, \\
\nn
\ee
where $\kappa = \alpha M_r$, and $C_m^1$ is a Gegenbauer polynomial.
Before calculating the spectrum
in this basis, in order to improve convergence, we apply a similarity 
transformation to the effective Hamiltonian in Eq.\ (\ref{schroedinger}), 
by means of which
\be
V ({\bf p}, {\bf p}') \longrightarrow 
\frac{1}{\sqrt{2 \omega^A_{\bf p} 2 \omega^B_{\bf p}}} \, 
V ({\bf p}, {\bf p}') \sqrt{2 \omega^A_{{\bf p}'} 2 \omega^B_{{\bf p}'}} \:,
\ee
while the kinetic term remains unchanged.

\begin{figure}[t]
\centerline{\includegraphics[width=12cm]{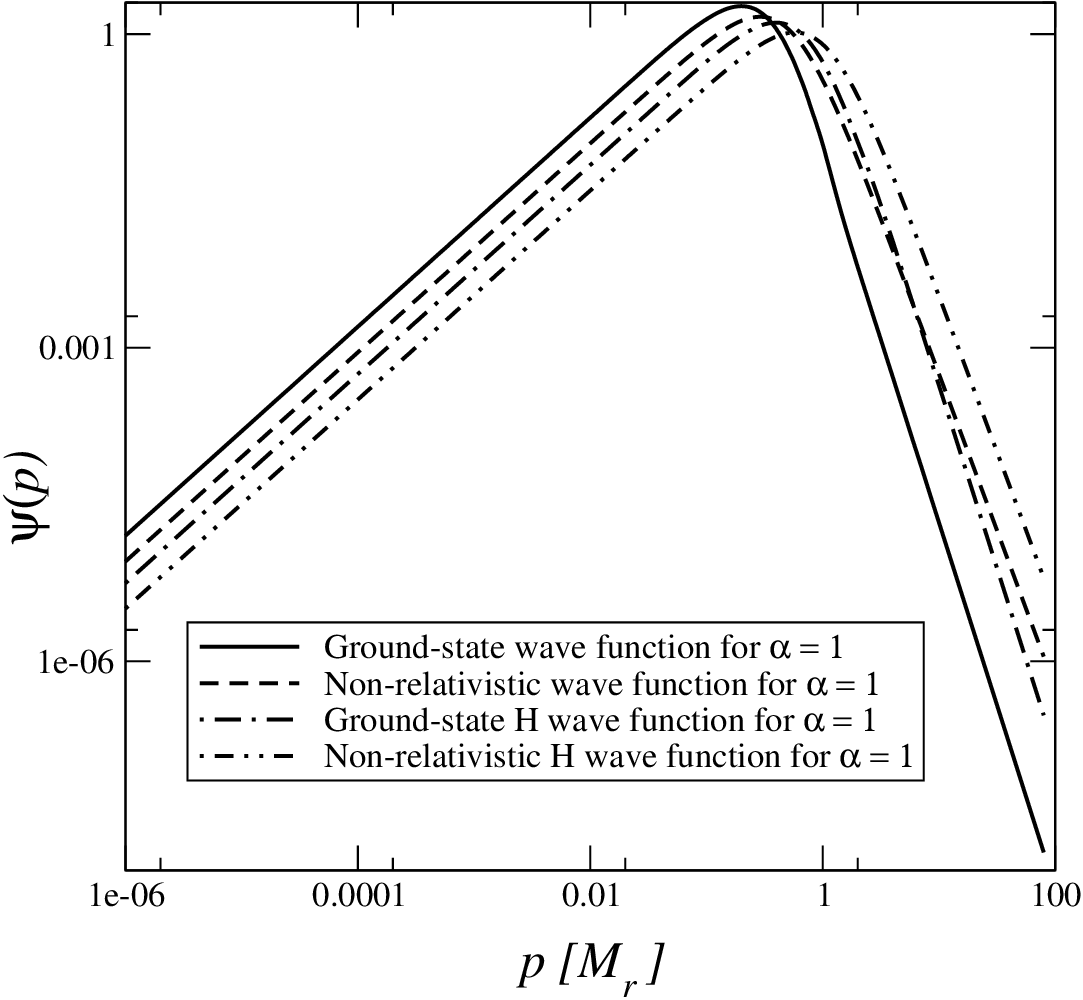}}
\caption{\label{fig.01} Double--logarithmic plot of the ground--state
wave function for $\alpha = 1$, compared to
the non--relativistic results. The H wave function refers to a mass ratio
$M_A/M_B$ as in the hydrogen atom, the other wave function is for equal 
masses.}
\end{figure}

\begin{figure}[t]
\centerline{\includegraphics[width=12cm]{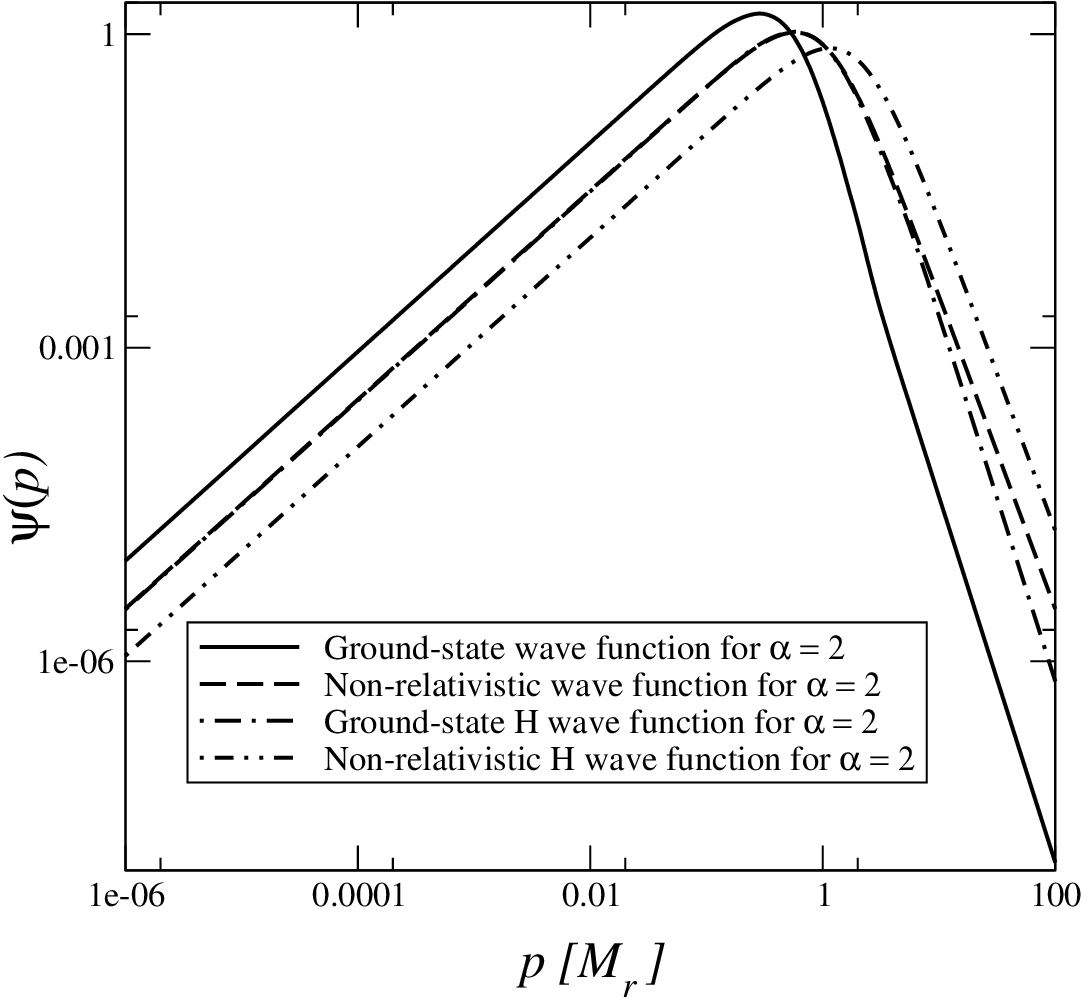}}
 \caption{\label{fig.02} Plot of the ground--state wave function as in
Fig.\ \ref{fig.01}, for $\alpha = 2$.}
\end{figure}

Away from the
weak--coupling limit, the eigenstates are quite different
from the corresponding Coulomb wave functions. Furthermore,
the Coulomb basis is not a good basis to describe these
states. The problem lies in the fact that all Coulomb
wave functions have the same high--momentum tail, and only their
low--momentum part varies.
Beyond the weak--coupling regime, the
high--momentum tail of the ground state of Eq.\ (\ref{schroedinger}) 
turns out to be 
suppressed with respect to the Coulomb wave functions (see Figs.\ 
\ref{fig.01} and \ref{fig.02}). 
Adjusting the overall scale of the 
Coulomb wave functions cannot cure this problem. 

Even though in the Coulomb basis
the energy converges rapidly with the number of basis
states, one can show that it does actually not converge to an eigenvalue
of Eq.\ (\ref{schroedinger}).
Therefore, we have used a different basis, with a dependence on the two
parameters $\beta$ and $\kappa$,
\be
\phi_n(p) = \frac{\sqrt{3 (2 n+1) (2 \beta - 1)}}
{\kappa^{3/2} \, [(p/\kappa)^3 + 1]^\beta} \,
P_n\left( 1 - \frac{2}{[(p/\kappa)^3 + 1]^{2 \beta -1}} \right) \:,
\ee
$P_n$ being a Legendre polynomial. This choice of basis leads to very
satisfying results over a wide range of intermediate values
of $\alpha$.  For most purposes the values $\kappa = \alpha M_r$
and $\beta = 3/2$ turned out to be sufficient.
The wave functions are normalized to unity,
\be
\int_0^\infty d p \, p^2 \phi_m(p) \phi_n(p) = \delta_{mn} \:.
\ee

We have tested our results by comparing $H_{BW} \psi (p)$ to
$E \psi (p)$ for the ground state, with the energy $E$ and wave function
$\psi (p)$ determined as described above. The deviations are typically
below one percent, which demonstrates that our numerical result for both the
energy and the wave function constitutes an excellent approximation. 
For $\alpha = 2$, the corresponding deviation in the case of
the Coulomb basis is of the order of 25\%. We have used up to 40 basis states, 
the changes in the deviation being minimal for more than 10 basis states. 
One can show that the difference
$(H_{BW} - E) \psi(p)$ is orthogonal to the basis with an
accuracy of $10^{-9}$ for our numerical integration, for both the Coulomb and
the other basis. This also demonstrates that the problem with the Coulomb 
basis has nothing to do with the numerics.

\begin{figure}[t]
\centerline{\includegraphics[width=10cm]{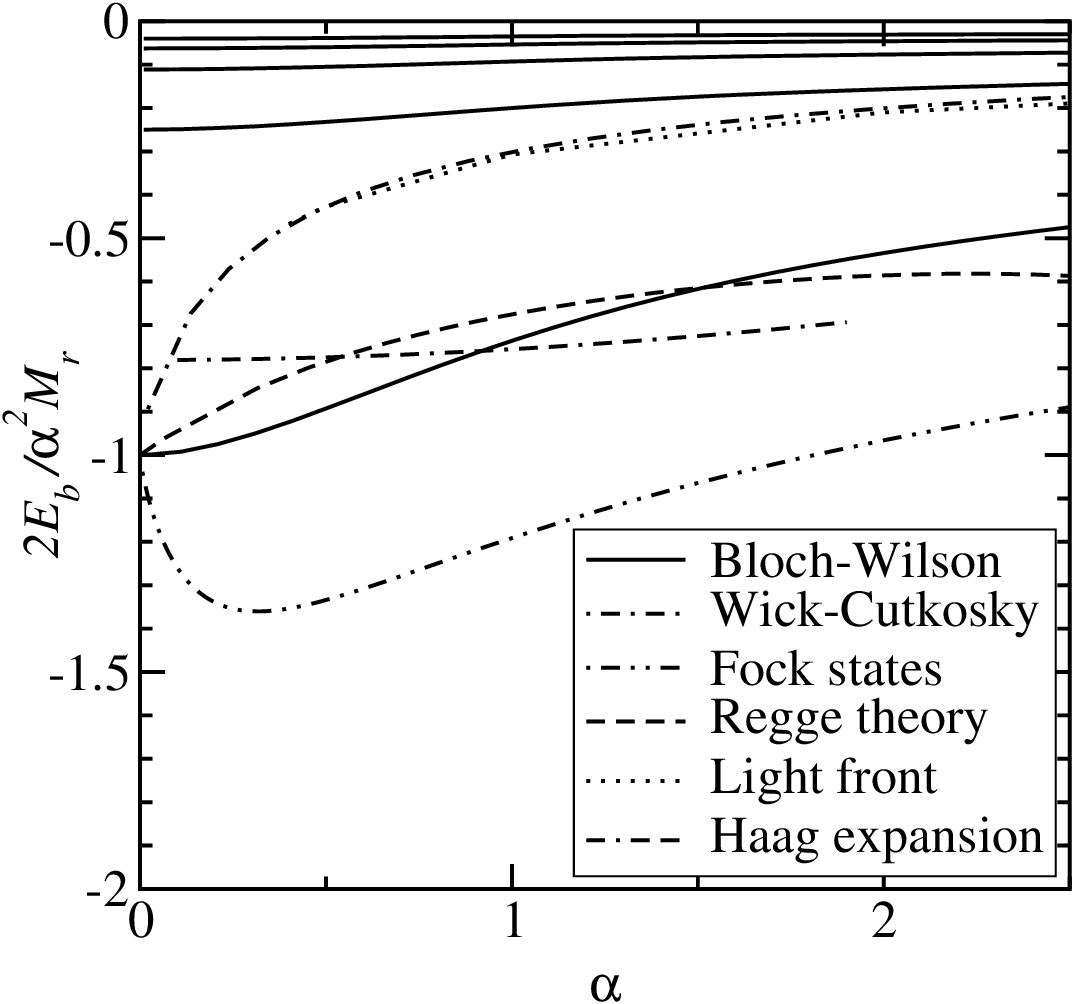}}
 \caption{\label{fig.03}  The spectrum of binding energies $E_b = E' - M_A
- M_B$ (cf.\ Eq.\ (\ref{schroedinger})) for $s$--states in
the equal--mass case, compared to the ground state energies of
the Wick-Cutkosky model \cite{Nak69}, the Hamiltonian eigenvalue 
equation in a Fock space truncation \cite{LB01}, the Regge theory predictions 
\cite{WLSD00}, the light--front calculation \cite{MC00}, and the Haag 
expansion results \cite{GRS95} in their domain of validity.}
\end{figure}

\begin{figure}[t]
\centerline{\includegraphics[width=10cm]{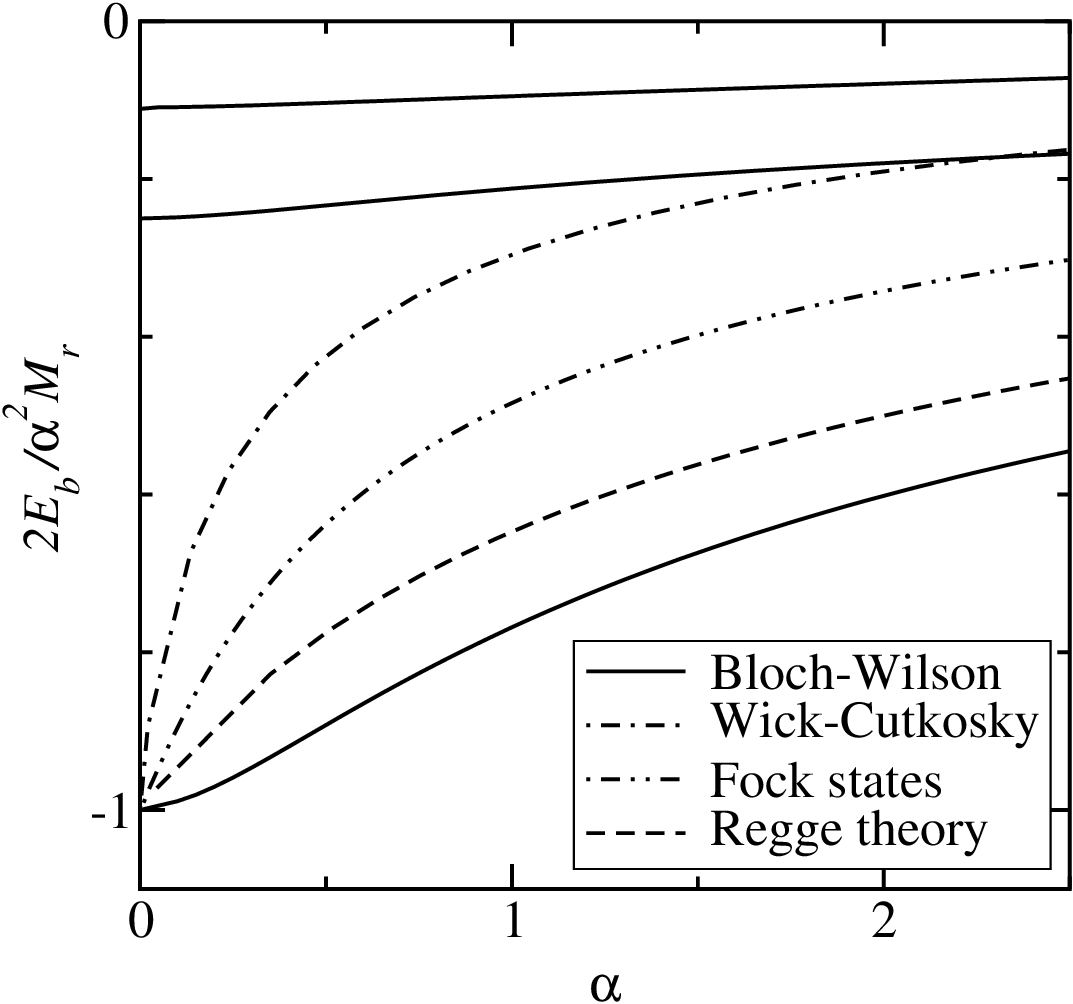}}
 \caption{\label{fig.04}  The spectrum of binding energies for $s$--states 
as in Fig.\ \ref{fig.03}, for the hydrogen--masses case.}
\end{figure}

The spectrum, as in most relativistic approaches, shows weaker
binding than is expected from the extrapolation of the 
non--relativistic formula (see Figs.\ \ref{fig.03} 
and \ref{fig.04}). 
Consequently,
the wave functions peak at a lower momentum. More interestingly,
the power--law behavior of the high--momentum tail of the
wave functions changes. Whilst the Coulomb
wave functions behave like $p \psi(p) \sim p^{-3}$,
the wave functions found here beyond the weak--coupling regime 
fall off more rapidly, with $p \psi(p) \sim p^{-3.5}$. The 
slope of the long--range tail in configuration space is unaltered
by the relativistic corrections (see Figs.\
\ref{fig.01} and \ref{fig.02}).

We have also made an attempt to visualize the non--local potential Eq.\
(\ref{relpot}). Given a solution $\psi({\bf p})$ of the Schr\"odinger 
equation Eq.\ (\ref{schroedinger}), one can define an ``effective'' local
potential $V_L ({\bf r})$ in position space by 
\be
V_L ({\bf r}) \psi ({\bf r}) &=& \left[ \int \frac{d^3 p'}{(2 \pi)^3} \, 
V({\bf p}, {\bf p}') \psi ({\bf p}') \right]_{F.T.} \nn \\[1mm]
&=& \left[ \left( E' - \sqrt{M_A^2 + {\bf p}^2} -
\sqrt{M_B^2 + {\bf p}^2} \right) \psi ({\bf p})\right]_{F.T.} \:,
\label{locpot}
\ee
where $F.T.$ denotes the Fourier transform to position space. The definition
Eq.\ (\ref{locpot}) makes sense at least for the ground state, given that the
latter has no nodes. Unfortunately, it is numerically very difficult to 
evaluate the expression Eq.\ (\ref{locpot}) over the whole range of $\bf r$. 
However, we could determine that, somewhat unexpectedly, at the origin 
${\bf r} = 0$, $V_L$ tends towards finite values, numerically 
$V_L (0) = -15.0$ for $\alpha = 1$ and $V_L (0) = -4.0$ for 
$\alpha = 2$.

\section{Discussion}

The Bloch--Wilson Hamiltonian has a number of practical advantages over 
other relativistic approaches. Most notable is the absence of
the energy eigenvalue in the potential part of the two--body 
bound--state equation. This
has important computational advantages over formulations that
have an energy--dependent integral kernel. 
We did not have to make any approximation for this result;
it is a natural consequence of our approach. In
the present formulation, the Hamiltonian is not Hermitian. Consequently, the 
left eigenstates are not identical to the right eigenstates, even though all 
eigenvalues turn out to be real.
As a consequence of the similarity transformation back to the $H_0$--invariant
subspace, the bound--state equation is expressed solely in terms of
the two--particle wave function; all the internal workings of the interaction
have gone in the Bloch--Wilson Hamiltonian. 

To the present order in the perturbative expansion, all formally
divergent contributions can be identified with proper on--shell Feynman
diagrams, hence the renormalization procedure used in covariant Lagrangian
perturbation theory can be applied to the effective Hamiltonian. This may
not be possible at higher orders, nor can it be expected, of course, 
when a non--perturbative approach for the determination of the 
Bloch--Wilson Hamiltonian is chosen. In these cases, other and possibly not 
manifestly covariant renormalization procedures will have to be used. However, 
none of this is necessary in the present case. In particular, to the order 
considered the self--energy corrections merely serve to renormalize the 
masses.  

Led by the results of the FSR approach \cite{NT96,STG99}, it is nowadays 
believed that the Bethe--Salpeter equation suffers,
apart from other unphysical features, also from a large underbinding. 
Our results lie closer to the non--relativistic values than the
Bethe--Salpeter results and most other relativistic approaches 
(see Figs.\ \ref{fig.03} and \ref{fig.04}), i.e., we find indeed a much 
stronger binding. Recently a discussion has emerged as to how 
the instability of the Wick--Cutkosky Hamiltonian could affect the bound state 
results \cite{GST01}--\cite{AA99}. At the present level of approximation, 
there seem to be no consequences of this unphysical feature of the model, 
however, how our approach can be extended to
be sensitive to the instability is a topic for future research.

Not only is the non--relativistic limit properly recovered, as we were able to 
demonstrate both analytically and numerically, but also the 
one--body limit \cite{Gro82}, where one particle becomes 
infinitely heavily, is consistent; the Schr\"odinger equation reduces to the
equation for one particle interacting with a fixed source. Therefore, unlike 
the Bethe--Salpeter equation, the present approach can be used to study 
heavy--light systems, such as the hydrogen atom. 

Finally, we comment on the invariance of the results under Lorentz
boosts. In principle, bound states in a moving frame, i.e., with total
momentum different from zero, can be calculated by solving the effective
Schr\"odinger equation Eq.\ (\ref{twopschroed}), and the results can be 
compared to the relativistic energy--momentum relation. 
At any rate, a perturbative Hamiltonian approach which by its very nature is 
not manifestly covariant, cannot be expected to maintain dynamical 
invariances, like boost invariance, exactly. As in the description
of any non--perturbative phenomenon, some symmetries of the underlying theory 
will be violated at any level of approximation. A bound state in a moving 
frame will then slightly differ from a bound state at rest. This difference 
will become smaller with increasing order in the coupling constant to which 
the Bloch--Wilson Hamiltonian is calculated. However, we emphasize that in
the present approach the expansion parameter is a Lorentz--invariant quantity,
and the necessary renormalization is carried through in a covariant way,
so that the violation of boost invariance might be expected to be small.
To what extent boost invariance is actually broken at the present 
order and how much of it can be restored by calculating the Hamiltonian to a 
higher order is a topic of current research.

\section*{Acknowledgements}

One of us (N.E.L.) would like to thank the Austrian Academy of Science and the
organizers of the  conference ``Quark Confinement and the Hadron Spectrum IV'' 
in Vienna, July 2000, for the financial support to attend the conference which
led to the present work.


\begin{thebibliography}{99}
\bibitem{GML51} M. Gell-Mann and F. Low, Phys.\ Rev.\ {84} (1951) 350.
\bibitem{FW71} A.L. Fetter and J.D. Walecka, Quantum Theory of
  Many--Particle Systems (McGraw--Hill, New York, 1971).
\bibitem{Web99} A. Weber, in Particles and Fields --- Seventh Mexican 
  Workshop, eds.\ A. Ayala, G. Con\-tre\-ras, and G. Herrera, AIP Conference 
  Proceedings 531 (American Institute of Phys\-ics, New York, 2000), 
  hep--th/9911198.
\bibitem{IZ80}C. Itzykson and J.--B. Zuber, Quantum Field Theory
(McGraw--Hill, New York, 1980).
\bibitem{BS51} E.E. Salpeter and H.A. Bethe, Phys.\ Rev.\ {84} (1951)
  1232.
\bibitem{Nak69} N. Nakanishi, Prog.\ Phys.\ (Suppl.) {43} (1969) 1;
  ibid.\ {95} (1988) 1.
\bibitem{Tod71} I.T. Todorov, Phys.\ Rev.\ D {3} (1971) 2351.
\bibitem{BS66} R. Blankenbecler and R. Sugar, Phys.\ Rev.\ {142}
  (1966) 1051.
\bibitem{LT63} A.A. Logunov and A.N. Tavkhelidze, Nuovo Cim.\ 29 (1963) 380.
\bibitem{Gro69} F. Gross, Phys.\ Rev.\ {186} (1969) 1448.
\bibitem{PW98} D.R. Phillips and S.J. Wallace, Few Body Syst.\ {24} (1998) 175.
\bibitem{ST93} Yu.A. Simonov and J.A. Tjon, Ann.\ Phys.\ (N.Y.) {228} (1993) 1.
\bibitem{Nie95} T. Nieuwenhuis, PhD thesis, University of Utrecht, 
The Netherlands, Oct.\ 1995.
\bibitem{NT96} T. Nieuwenhuis and J.A. Tjon, Phys.\ Rev.\ Lett.\ {77} (1996) 
  814.
\bibitem{BPP98} S.J. Brodsky, H.-C. Pauli, and S.S. Pinsky, Phys.\ Rep.\ 
  {301} (1998) 299.
\bibitem{CDKM98} J. Carbonell, B. Desplanques, V.A. Karmanov, and J.F. Mathiot,
  Phys.\ Rep.\ 300 (1998) 215.
\bibitem{BHM99} S.J. Brodsky, J.R. Hiller, and G. McCartor, Phys.\ Rev.\ 
  {D 60} (1999) 054506.
\bibitem{Per94}R.J. Perry, Ann.\ Phys.\ (N.Y.) 232 (1994) 116.
\bibitem{GW98}E.L. Gubankova and F. Wegner, Phys.\ Rev.\ D 58 (1998) 025012.
\bibitem{Wal99} T.S. Walhout, Phys.\ Rev.\ D 59 (1999) 065009.
\bibitem{Ji94}C.R. Ji, Phys.\ Lett.\ {B 322} (1994) 389.
\bibitem{MC00} M. Mangin-Brinet and J. Carbonell, Phys.\ Lett.\ {B 474} (2000) 
  237.
\bibitem{Han00} J. Hansper, Phys.\ Rev.\ D 62 (2000) 056001.
\bibitem{GRS95} O.W. Greenberg, R. Ray, and F. Schlumpf, Phys.\ Lett.\ 
  {B 353} (1995) 284.
\bibitem{LB01} N.E. Ligterink and B.L.G. Bakker, hep--ph/0010167.
\bibitem{WLSD00} A. Weber, J.C. L\'opez Vieyra, C.R. Stephens, S. Dilcher, 
  and P.O. Hess, hep--th/0011280.
\bibitem{WC54} G.C. Wick, Phys.\ Rev.\ {96} (1954) 1124;
  R.E. Cutkosky, Phys.\ Rev.\ {96} (1954) 1135.
\bibitem{Gol57} J. Goldstone, Proc.\ Roy.\ Soc.\ (London) {A239} (1957) 267.
\bibitem{BS57} H.A. Bethe and E.E. Salpeter, Quantum  mechanics of one--
  and two--electron atoms (Springer, Berlin, 1957).
\bibitem{Gro82} F. Gross, Phys.\ Rev.\ C {26} (1982) 2203.
\bibitem{STG99} C. Savkli, J.A. Tjon, and F. Gross,
  Phys.\ Rev.\ C 60 (1999) 055210; Erratum ibid.\ C 61 (2000) 069901.
\bibitem{GST01} F. Gross, C. Savkli, and J. Tjon, nucl--th/0102041.
\bibitem{RS96} R. Rosenfelder and A.W. Schreiber, Phys.\ Rev.\ D 53 (1996) 
  3337, 3354.
\bibitem{AA99} S. Ahlig and R. Alkofer, Ann.\ Phys.\ (N.Y.) 275 (1999) 113.
\end{thebibliography}
\end{document}